\documentclass[article,preprint,12pt,unsrt]{elsarticle}
\usepackage{a4wide}
\usepackage{natbib} 			
\usepackage{graphicx}
\usepackage{amsfonts}
\usepackage{amsmath}
\usepackage{booktabs}
\usepackage{epsfig}
\usepackage{stmaryrd}
\usepackage{hyperref}
\hypersetup{citecolor=black,linkcolor= black}

\begin{document}
\title{Power-law correlations in finance-related Google searches, and their cross-correlations with volatility and traded volume: Evidence from the Dow Jones Industrial components}
\author{Ladislav Kristoufek}
\ead{kristoufek@icloud.com}
\address{
Institute of Information Theory and Automation, Czech Academy of Sciences, Pod Vodarenskou Vezi 4, 182 08, Prague 8, Czech Republic\\
Warwick Business School, University of Warwick, Coventry, West Midlands, CV4 7AL, United Kingdom
}

\begin{abstract}
We study power-law correlations properties of the Google search queries for Dow Jones Industrial Average (DJIA) component stocks. Examining the daily data of the searched terms with a combination of the rescaled range and rescaled variance tests together with the detrended fluctuation analysis, we show that the searches are in fact power-law correlated with Hurst exponents between 0.8 and 1.1. The general interest in the DJIA stocks is thus strongly persistent. We further reinvestigate the cross-correlation structure between the searches, traded volume and volatility of the component stocks using the detrended cross-correlation and detrending moving-average cross-correlation coefficients. Contrary to the universal power-law correlations structure of the related Google searches, the results suggest that there is no universal relationship between the online search queries and the analyzed financial measures. Even though we confirm positive correlation for a majority of pairs, there are several pairs with insignificant or even negative correlations. In addition, the correlations vary quite strongly across scales.
\end{abstract}


\journal{Physica A}

\maketitle


\section{Introduction}

Analysis of online activity of the internet users has proved its worth in various disciplines, most notably in psychology \cite{McCarthy2010,Page2011,Sueki2011,Yang2011,Hagihara2012}, ecology \cite{McCallum2013,McCallum2014,Ficetola2013,Verissimo2014}, epidemiology \cite{polgreen2008,ginsberg2009,Carneiro2009,Seifter2010,Dugas2012}, medicine \cite{Linkov2014,Telem2014} linguistics \cite{Mocanu2013}, politology \cite{Metaxas2012}, sociology \cite{Rybski2009,Grabowicz2014} and in a wide range of economics, marketing and finance \cite{Preis2010,mondria2010,Goel2010,Vosen2011,Drake2012,Bank2011,Bordino2012,Vlastakis2012,Dzielinski2012,Preis2012a,Preis2013,Kristoufek2013,Kristoufek2013a,Moat2013,Curme2014}. In the economic and financial applications, the focus has been primarily put on the search queries on various search engines such as Google, Yahoo! and Baidu. Bank \textit{et al.} \cite{Bank2011} find connection between Google searches and liquidity at the German stock market. Bordino \textit{et al.} \cite{Bordino2012} study traded volume of the NASDAQ-100 index component stocks and they report that it is correlated with the related searches of the Yahoo! engine. Vlastakis \& Markellos \cite{Vlastakis2012} find positive correlation between internet search queries for NASDAQ and NYSE stocks and their traded volume and volatility. Dzielinski \cite{Dzielinski2012} introduces an uncertainty measure based on the financial online search queries. Preis \textit{et al.} \cite{Preis2013} show that Google searches for financial terms can be used for profitable trading strategies. Kristoufek \cite{Kristoufek2013} utilizes popularity of the Dow Jones stocks measured by Google search queries for portfolio diversification. Kristoufek \cite{Kristoufek2013a} further studies dynamics between Google searches, Wikipedia page views and dynamics of the Bitcoin crypto-currency uncovering a strong relationship between these. Moat \textit{et al.} \cite{Moat2013} report that even Wikipedia page views can be utilized for the trading strategy construction. And Curme \textit{et al.} \cite{Curme2014} cluster the online searches into groups and show that mainly politics and business oriented searches are connected to the stock market movements.

The most frequently reported relationship between the online searches, traded volume and volatility directs further to the dynamic characteristics of the online searches time series. As traded volume and volatility have been repeatedly studied for their power-law correlation structures \cite{Cont2001,Poon2003,Poon2005}, the same research line is at hand for the online searches as well. Potential long-term memory of the online activity has further implications for modeling and correct inspection of dynamics between the searches and other series. Here, we examine the correlation structure of the Google searches related to the Dow Jones Industrial Average (DJIA) index components. Daily Google searches data are utilized for the components of DJIA and as such, we present the first such study of the correlation structure of the online searches. To do so, we apply the rescaled range and rescaled variance tests to uncover the power-law correlations structure and we further proceed with the detrended fluctuation analysis of the search queries series. As it turns out that the DJIA-related Google queries are in fact power-law correlated, we reinvestigate a popular topic of cross-correlations between the searches, traded volume and volatility of the examined stocks. As we find the online searches to be power-law correlated and on the edge of (non)stationarity, we utilize the newly proposed correlation coefficients based on the detrended cross-correlation and detrending moving-average cross-correlation analyses.

The paper is organized as follows. In Section 2, we describe the used methodology, specifically the rescaled range and rescaled variance tests together with the moving block bootstrap significance criterion, and the detrended fluctuation analysis as well as the correlation coefficients. Section 3 introduces the dataset and presents the results. Section 4 concludes. We show that the Google searches related to the DJIA component stocks show scaling characteristic for the power-law correlated processes. This is supported by all utilized methods. General interest in the publicly traded companies thus shares similar properties to the variance and traded volume series -- there are profound periods of high interest followed by long-lived periods of low interest. However, the search series always revert back to a long-term trend so that no explosive behavior is observed. After taking the long-term memory aspect of the online query series into consideration, the correlations between the searches, traded volume and volatility become quite unstable and no universal relationship is found. The initial long-term memory analysis thus proves to be crucial for a correct treatment of cross-correlations between the online searches and various possibly connected series.

\section{Methodology}

\subsection{Long-term memory and its tests}

Long-term memory (or alternatively long-range dependence and long-range correlations) is defined through a power-law decay of the auto-correlation function $\rho(k)$ which scales as $\rho(k) \propto k^{2H-2}$ for lag $k \rightarrow +\infty$ \cite{Hurst1951,Beran1994,Samorodnitsky2006}. The series are then referred to as the power-law (auto-)correlated processes as well. The characteristic parameter of such processes is Hurst exponent $H$, or alternatively parameter $\alpha$, which takes values between 0 and 1 for stationary processes. The breaking value of $H=0.5$ characterizes a process with no long-term memory. Processes with $H>0.5$ are usually referred to as persistent processes whereas the ones with $H<0.5$ as anti-persistent processes. The former ones are reminiscent of locally trending processes which, however, keep their stationarity (for $H<1$) and return to their mean value quickly enough. The latter ones are very erratic in behavior as they switch their direction more frequently than uncorrelated processes. Integrating the stationary long-range dependent processes once creates an additional category of processes which have interesting properties. For $1 \le H < 1.5$, we have non-stationary yet still mean reverting processes. The frontier of $H=1.5$ marks a unit root process and $H>1.5$ characterizes processes which are non-stationary and not mean reverting, i.e. explosive processes. The long-term memory property of time series has far-reaching consequences for the time series modeling and forecasting mainly due to its implication of a non-summable auto-correlation function \cite{Samorodnitsky2006}. Therefore, it is essential to distinguish between long-range dependence with its power-law correlations and short-range dependence with its exponential correlation structure. For this purpose, we utilize the modified rescaled range test and the rescaled variance test.

The modified rescaled range test \cite{Lo1991} is an adjusted version of the original rescaled range analysis \cite{Hurst1951}. Both methods are based on scaling of the rescaled ranges with an increasing time series length. For the time series $\{x_t\}$ with $t=1,2,\ldots,T$, the testing statistic $V_T$ is defined as
\begin{equation}
\label{MRS}
V_{T}=\frac{R}{S\sqrt{T}} \nonumber
\end{equation}
where $R$ is a range of the profile of the analyzed series, 
\begin{equation}
R=\max_{t=1,\ldots,T}\left(\sum_{i=1}^{t}{(x_i-\bar{x}})\right)-\min_{t=1,\ldots,T}\left(\sum_{i=1}^{t}{(x_i-\bar{x}})\right), \nonumber
\end{equation}
with $\bar{x}$ being the time series average, $S$ is a heteroskedasticity and autocorrelation consistent (HAC) estimator of the standard deviation of the original series, defined as
\begin{equation}
\label{S}
S^2=\widehat{\gamma}(0)+2\sum_{k=1}^q{\left(1-\frac{k}{q+1}\right)\widehat{\gamma}(k)},
\end{equation}
with $\widehat{\gamma}(k)$ being an estimated auto-correlation with lag $k$ using the Barlett-kernel weights. Note that $\widehat{\gamma}(0)$ is an estimated variance. The crucial difference between the original and the modified version of the test stems in Eq. \ref{S} which is constructed to control for a possible short-term memory bias. Selection of the parameter $q$ then becomes crucial as an overshot parameter $q$ can suppress even long-term memory whereas an undershot $q$ parameter can direct to a misleadingly found long-term memory which in fact is only a strong short-term memory. We stick to an automatic selection criterion of the parameter as proposed by Lo \cite{Lo1991}
\begin{equation}
\label{eq2}
q^{\ast}=\left\lfloor\left(\frac{3T}{2}\right)^{\frac{1}{3}}\left(\frac{2|\widehat{\rho}(1)|}{1-\widehat{\rho}(1)^2}\right)^{\frac{2}{3}}\right\rfloor
\end{equation}
where $\widehat{\rho}(1)$ is the sample first order autocorrelation and $\lfloor \rfloor$ is the lower integer operator.

The rescaled variance test \cite{Giraitis2003} is based on a very similar idea as the previous one but, as the name suggests, it is based on the profile variance rather than the profile range so that it is less sensitive to extreme values. The testing statistic $M_T$ is then defined as
\begin{equation}
M_T=\frac{\text{var}(X)}{TS^2} \nonumber
\end{equation}
where $\text{var}(X)$ is the variance of the profile of the original series. To control for the short-term memory bias, the HAC standard deviation from Eq. \ref{S} and the optimal $q^{\ast}$ from Eq. \ref{eq2} are used here as well.

Even though both $V_T$ and $M_T$ have well defined asymptotic critical values \cite{Lo1991,Giraitis2003}, we opt for an alternative approach utilizing the moving block bootstrap methodology \cite{Efron1993,Srinivas2000} due to a finite sample, a very heterogenous dynamics of the analyzed series as well as their distributional properties. In the procedure, surrogate series are formed by shuffling the blocks of a fixed size from the original series. This way, the short-term correlations and distributional properties are kept but the long-term correlations are shuffled away creating a distribution of the testing statistic under a more realistic null hypothesis. In our application, we fix the block size to 25 observations and we bootstrap 1000 surrogate series to obtain statistical significance. 

\subsection{Detrended fluctuation analysis}

Detrended fluctuation analysis (DFA) \cite{Peng1993,Peng1994,Kantelhardt2002} is the most popular and the most frequently applied time domain estimator of Hurst exponent. This is mainly due to the fact that DFA works under various settings such as non-stationarity and trends \cite{Kantelhardt2002}, periodic cycles and seasonalities \cite{Hu2001}, and heavy tails \cite{Barunik2010}.

The procedure is based on the following steps. We work with the profile $X(t)$ of the series $\{x_t\}$ with $t=1,\ldots,T$ defined as
\begin{equation}
X(t)=\sum_{i=1}^{t}{(x_i-\bar{x})}. \nonumber
\end{equation}
The profile is divided into $T_s\equiv \lfloor T/s \rfloor$ non-overlapping windows of length $s$ which is referred to as a scale. The time series length $T$ may be non-divisible by $s$ which creates an issue with the end of the series which would not be used in the procedure. For this purpose, the series is in addition divided into boxes from the end of the series so that we obtain $2T_s$ boxes of size $s$. In each of these boxes, we calculate a mean squared deviation from the linear time trend inside the box. This means that for the $j$th box of size $s$, we obtain 
\begin{equation}
F^2(j,s)=\frac{1}{s}\sum_{i=1}^{s}{(X(s[j-1]+i)-\widehat{X_j(i)})^2} \nonumber
\end{equation}
where $\widehat{X_j(i)}$ is a linear fit of a time trend at position $i$ in window $j$. In a similar manner, we obtain the fluctuation for the boxes formed from the end of the series as
\begin{equation}
F^2(j,s)=\frac{1}{s}\sum_{i=1}^{s}{(X(T-s[j-T_s]+i)-\widehat{X_j(i)})^2}. \nonumber
\end{equation}
We then construct a fluctuation for specific scale $s$ as
\begin{equation}
F(s)=\left(\frac{1}{2T_s}\sum_{j=1}^{2T_s}{[F^2(j,s)]} \right)^{\frac{1}{2}} \nonumber
\end{equation}
and finally, we obtain Hurst exponent via the scaling law
\begin{equation}
\label{HH}
F(s) \propto s^H.
\end{equation}
In the application, we estimate the exponent for scales between $s_{min}=10$ and $s_{max}=500 \approx T/5$. Moreover for better illustrational purposes, we base the estimation and the results on scales $s$ which are powers of 10 to a single decimal point. Note, however, that the results do not change qualitatively for other specifications of scales and box splitting procedures and such approach is thus kept primarily for a straightforward presentation of the results.

\subsection{DCCA and DMCA coefficients}

The detrended cross-correlation coefficient $\rho_{DCCA}(s)$ for scale $s$ as proposed by Zebende \cite{Zebende2011} is a combination of the detrended fluctuation analysis (DFA) \citep{Peng1993,Peng1994,Kantelhardt2002} and the detrended cross-correlation analysis (DCCA) \citep{Podobnik2008,Zhou2008,Jiang2011}. The coefficient is defined as
\begin{equation}
\rho_{DCCA}(s)=\frac{F^2_{DCCA}(s)}{F_{DFA,x}(s)F_{DFA,y}(s)}, \nonumber
\label{rho}
\end{equation}
where $F^2_{DCCA}(s)$ is a detrended covariance between profiles of series $\{x_t\}$ and $\{y_t\}$ based on a window of size $s$, and $F^2_{DFA,x}$ and $F^2_{DFA,y}$ are detrended variances of profiles of the separate series, respectively, for a window size $s$\footnote{DCCA is a bivariate generalization of DFA presented in the previous section.}. For time series of length $T$, the series is divided into non-overlapping boxes of length $s$. In each box, fluctuation functions are computed for linearly detrended series which are in turn averaged over all boxes of the same length. In the case when $T$ is not divisible by $s$, the series is divided from the beginning as well as from the end and the averages are based on these sub-periods as in the case of DFA. More details about the methods and some alternative specifications can be found in Refs. \cite{Kantelhardt2002,Podobnik2008,Kristoufek2014,He2011,He2011b,Cao2014,Wang2014}. 

The detrending moving-average cross-correlation coefficient $\rho_{DMCA}(\lambda)$ for scale $\lambda$ has been introduced by Kristoufek \cite{Kristoufek2014a} as an alternative to the DCCA coefficient. The method builds on a connection between the detrending moving average (DMA) procedure \citep{Vandewalle1998,Alessio2002} and detrending moving-average cross-correlation analysis (DMCA) \citep{Arianos2009,He2011a}. The coefficient is defined as
\begin{equation} 
\rho_{DMCA}(\lambda)=\frac{F_{DMCA}^2(\lambda)}{F_{x,DMA}(\lambda)F_{y,DMA}(\lambda)}, \nonumber
\end{equation}
where $F^2_{DMCA}(\lambda)$, $F^2_{DMA,x}(\lambda)$ and $F^2_{DMA,y}(\lambda)$ are a detrended covariance between profiles of the examined series and detrended variances of the separate series, respectively, with a moving average parameter $\lambda$. Fluctuation functions are based on series detrended by a centered moving average of length $\lambda$. Various specifications can be utilized for the detrending but the centered averaging has been shown to outperform the contenders \cite{Carbone2003}. Contrary to the DCCA coefficient, the DMCA coefficient is not based on a box-splitting and it is thus computationally more efficient. More details can be found in Refs. \cite{Kristoufek2014a,Alessio2002,Arianos2009}.

In a series of papers, Kristoufek \cite{Kristoufek2014,Kristoufek2014a} shows that the statistical properties of both methods depend strongly on long-term memory properties of the separate series. Moreover, reliability of the methods is not constant for different levels of correlation between the studied processes either. To control for such effects, we apply the Theiler's Amplitude Adjusted Fourier Transform (TAAF) \citep{Theiler1992}. This method reconstructs the series with the same spectral as well as distributional properties as the original one. This way, we obtain two series with an unchanged auto-correlation and distributional structure which are, however, pairwise uncorrelated. Statistical significance of estimated correlations based on DCCA and DMCA can be then obtained and tested.

Specifically for each studied pair of processes, we obtain TAAF transformed series which are not cross-correlated but retain the auto-correlation and distribution properties of the original series. The DCCA and DMCA coefficients are then estimated for such series. As the series are not cross-correlated, the expected value of the coefficients is zero. However, variance of the estimates can be possibly high. Therefore, we estimate the coefficients on 1000 surrogate series to obtain a finite sample distribution under the null hypothesis of no cross-correlations between series which controls for both long-term auto-correlations and distributional properties.

\section{Data and results}

Google provides search query time series for specified terms from the year of 2004 onwards. However, the series are not reported as a pure number of searches for a given term but these are renormalized according to the Google algorithm which can be in essence seen as rescaling the searches into the 0--100 interval so that the number represents the proportion of the specified searched term among all searched terms in time being kept between 0 and 100. Moreover, the obtained numbers are based on sampling from all searched terms so that these represent an estimated rescaled proportion. Even though such rescaling procedure can somewhat dilute the information content of the series, the empirical results summarized in the introductory section show otherwise.

The Google data can be downloaded freely from the Google Trends website (trends.google.com) at a weekly frequency. To obtain the data at a higher frequency, specifically the daily one, one needs to download the series in three-months sections and the series further need to be rescaled and chained together. We apply such procedure for the component stocks of the Dow Jones Industrial Average (DJIA) index between years 2004 and 2013 (apart from Exxon Mobil, J. P. Morgan and Procter \& Gamble for which the series are several months shorter which will be evident later in the text) and thus obtaining 2516 observations for most series. The most severe issue with the Google queries data is its relative arbitrariness in defining the searched terms. Further, the sampling and thresholding procedure applied by Google for its search series quite frequently ends up with reporting incomplete series. If the specified term is not searched for frequently enough, the series is practically useless. We thus analyze only the component stocks which provide reliable search query series. Out of 30 DJIA stocks, we end up with 18 stocks for which the Google series are reliable without discontinuities. The analyzed stocks are summarized in Tab \ref{query}. We have tried various combinations and specifications of the searched terms and we report the ones which provided the most complete series.

\begin{figure}[!htbp]
\begin{center}
\begin{tabular}{ccc}
\includegraphics[width=50mm]{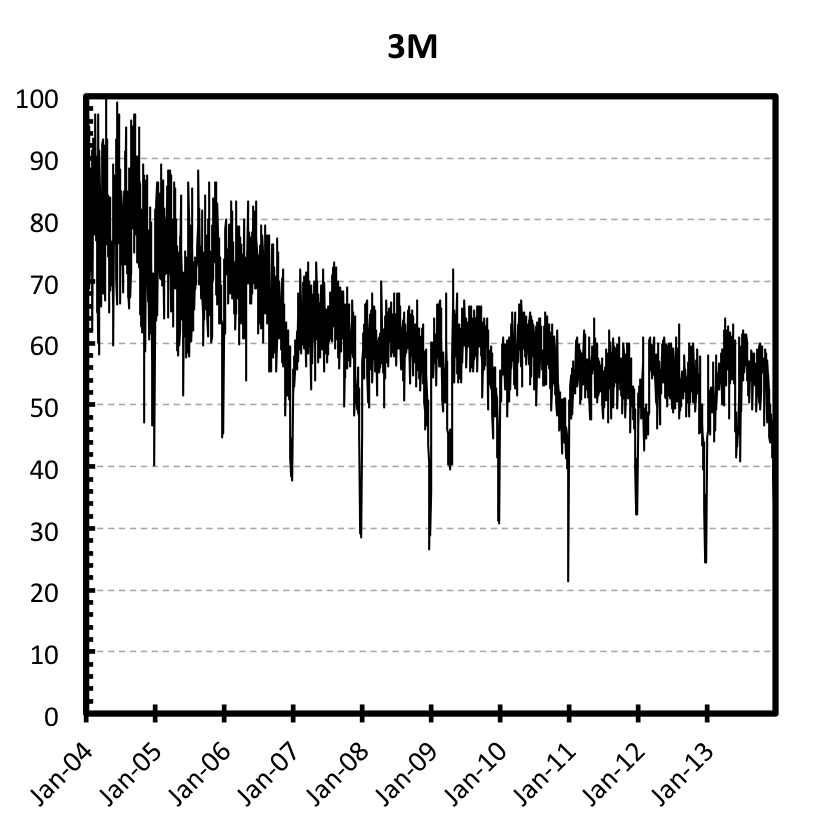}&\includegraphics[width=50mm]{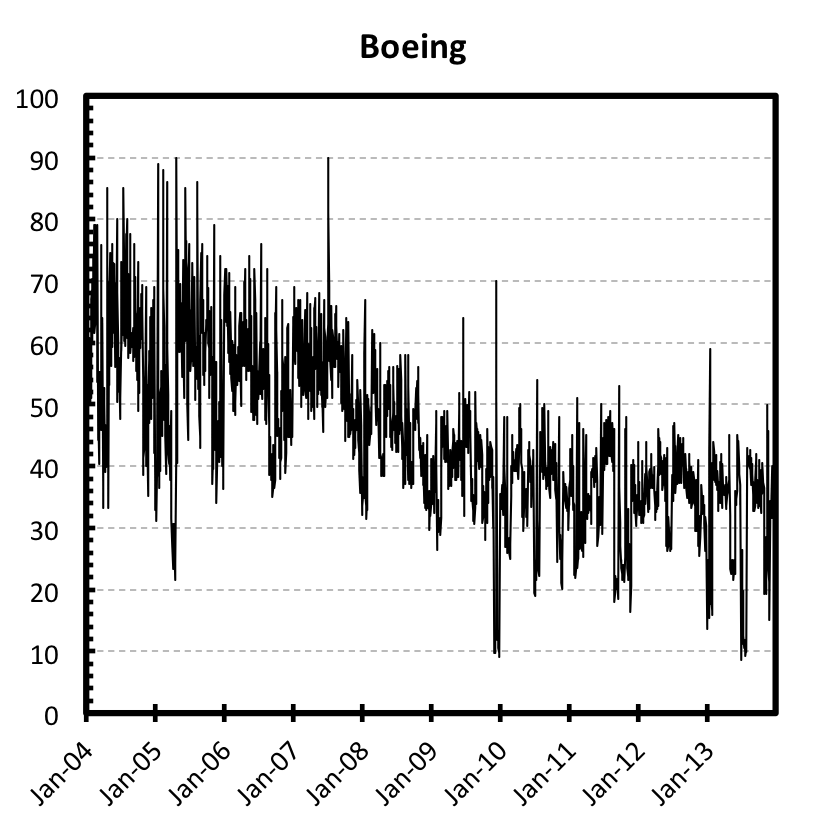}&\includegraphics[width=50mm]{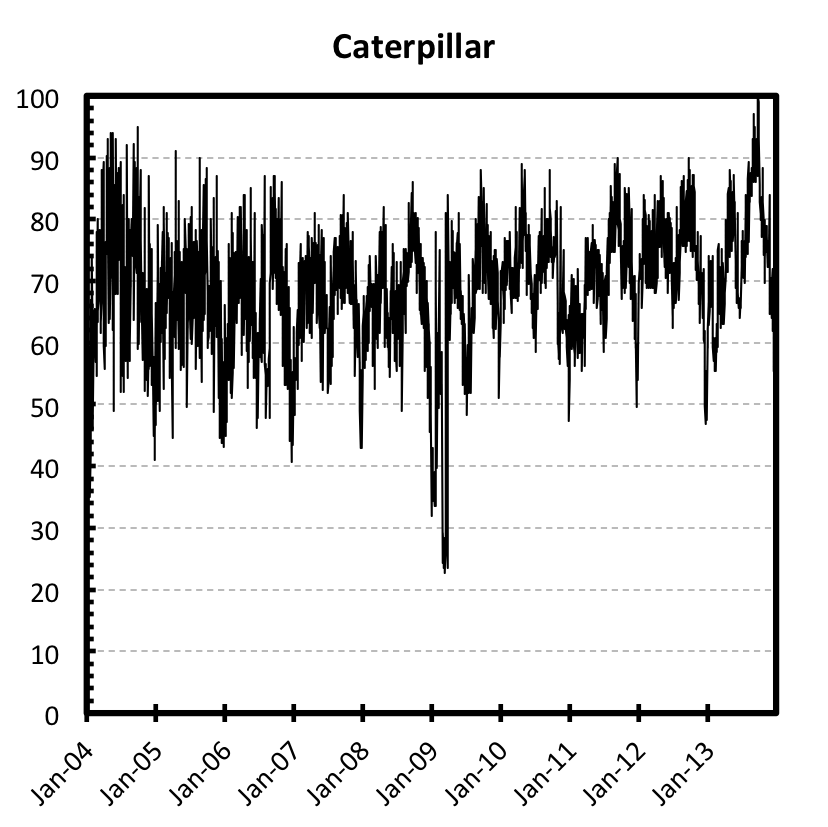}\\
\includegraphics[width=50mm]{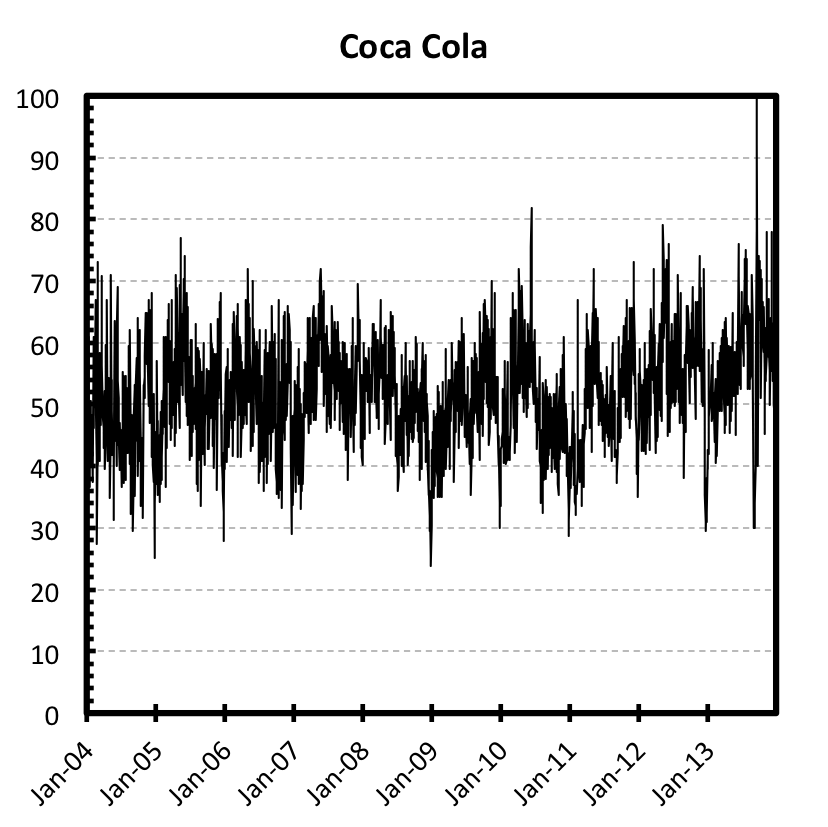}&\includegraphics[width=50mm]{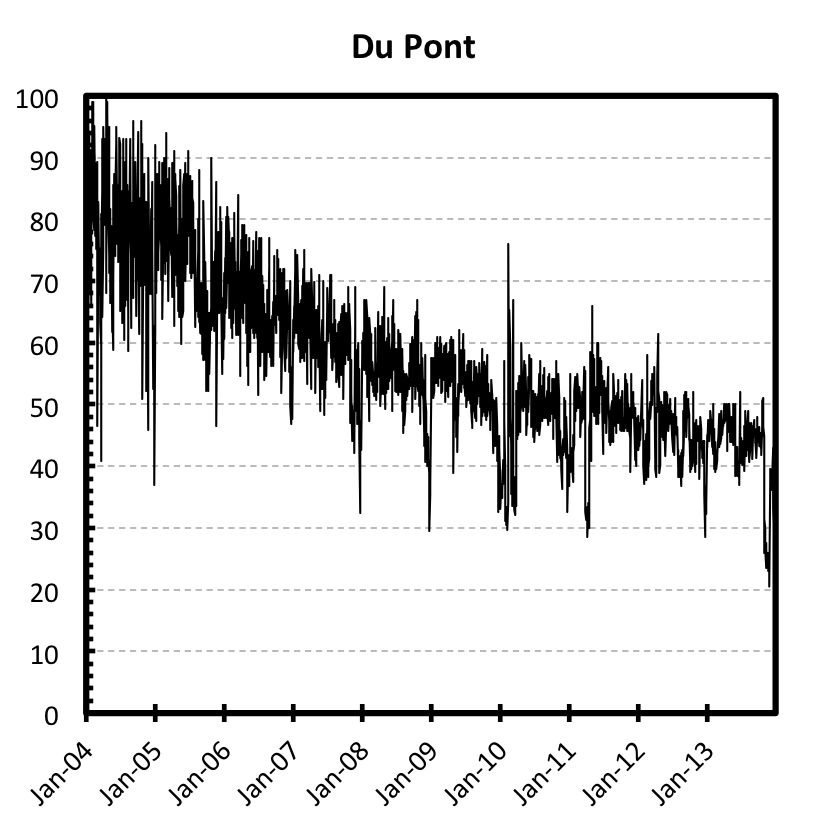}&\includegraphics[width=50mm]{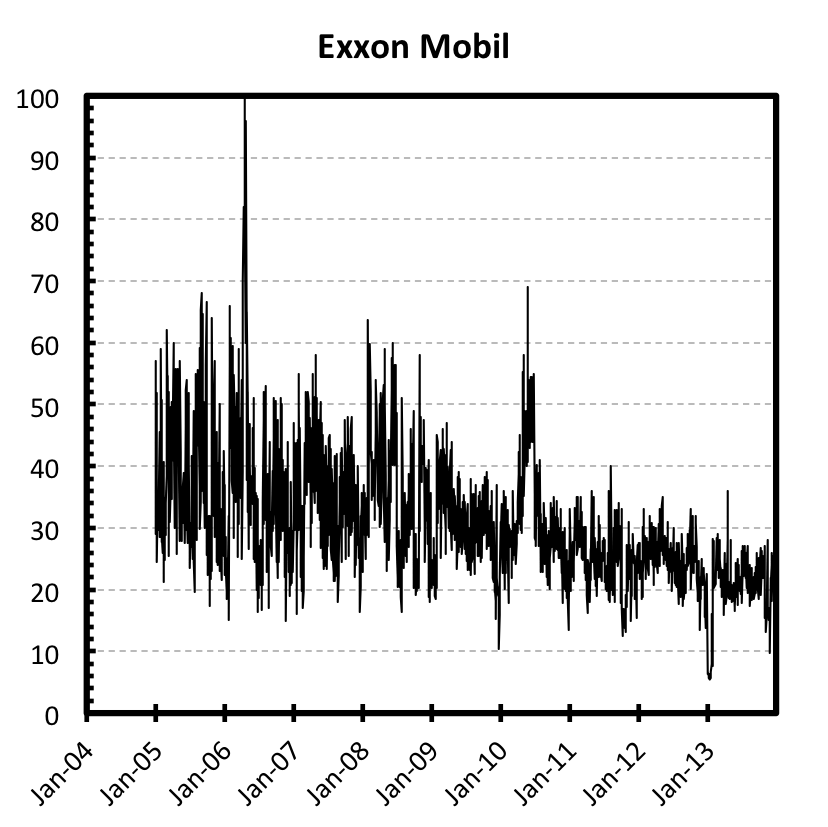}\\
\includegraphics[width=50mm]{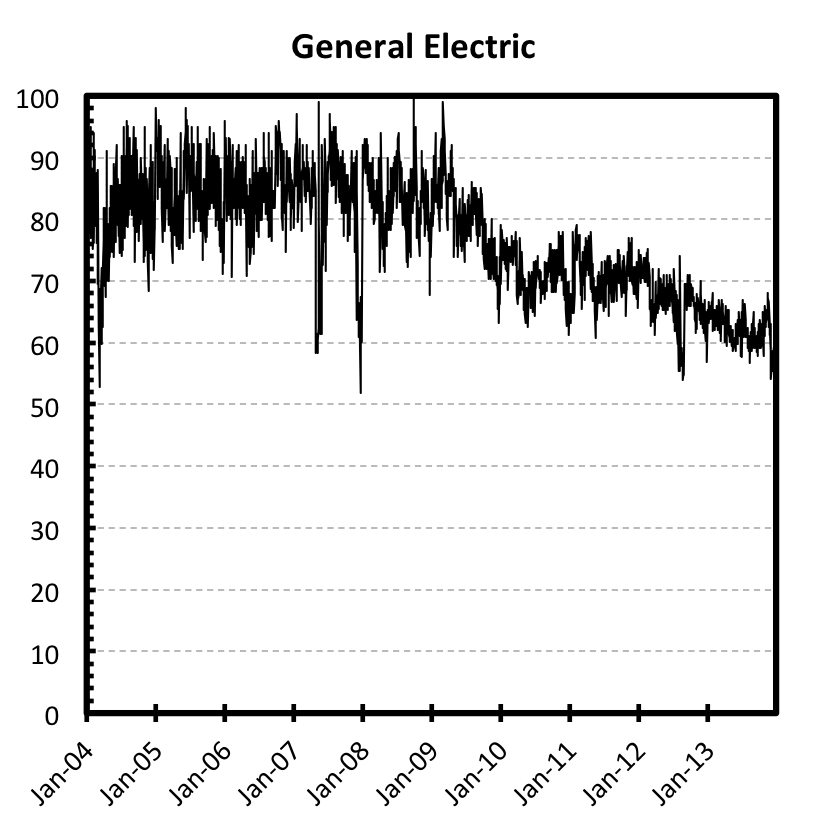}&\includegraphics[width=50mm]{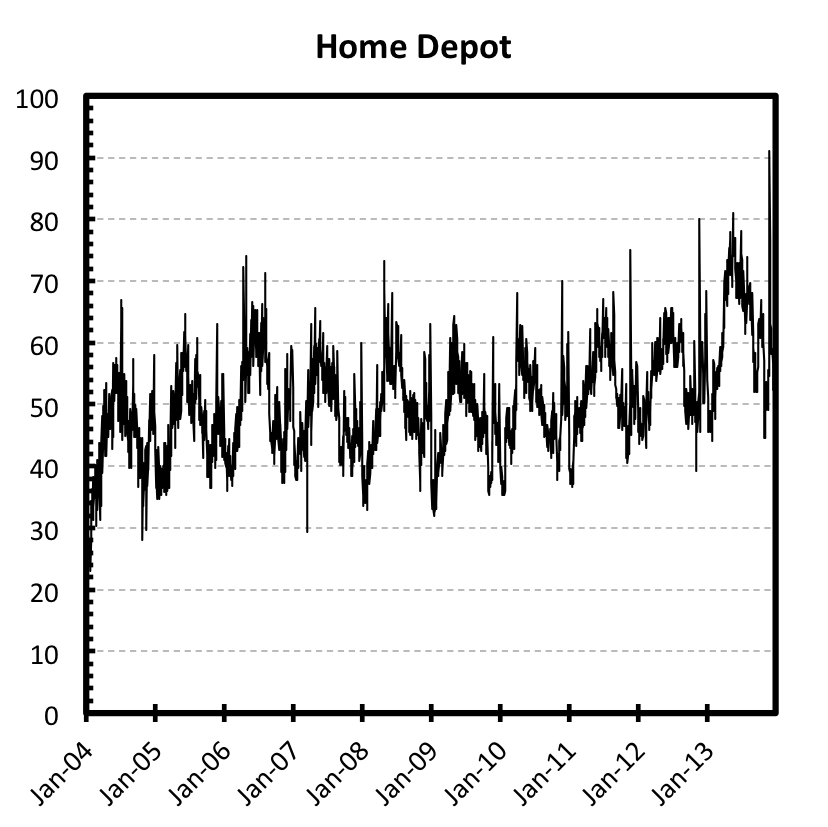}&\includegraphics[width=50mm]{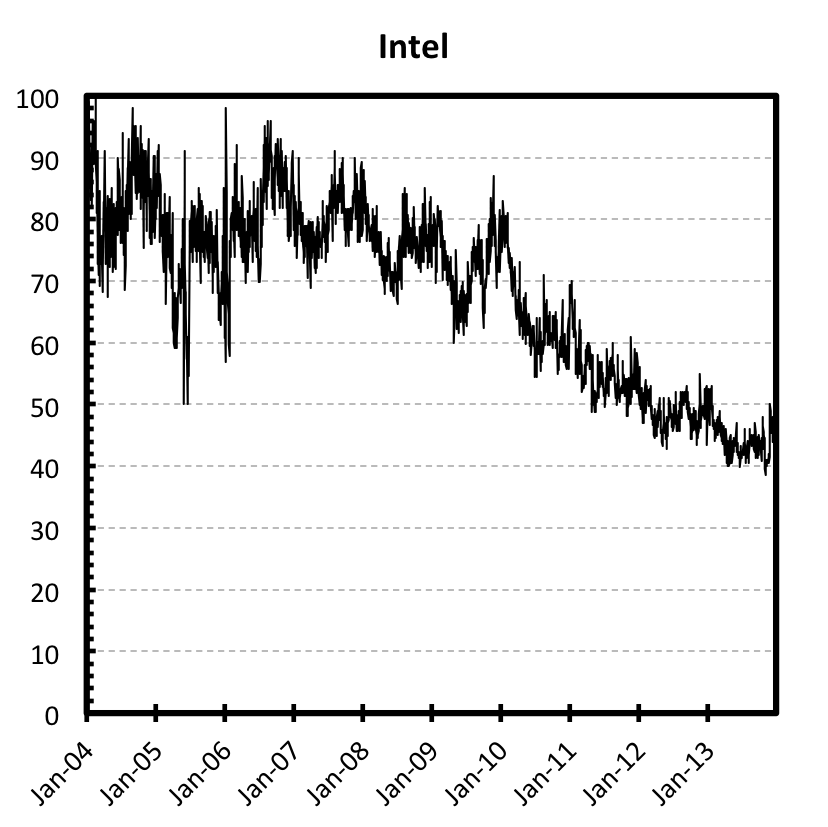}\\
\end{tabular}
\end{center}\vspace{-0.5cm}
\caption{\textbf{Normalized Google searches (Part 1).} \footnotesize{Covered period ranges between 1.1.2004 and 31.12.2013 with a daily frequency.}\label{fig_series1}
}
\end{figure}

\begin{figure}[!htbp]
\begin{center}
\begin{tabular}{ccc}
\includegraphics[width=50mm]{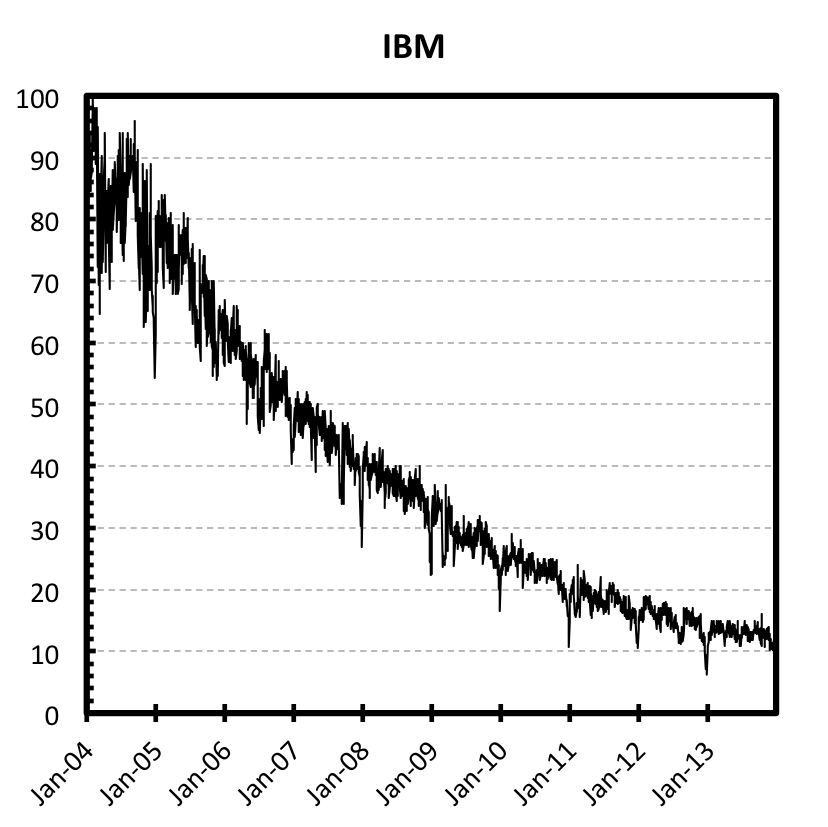}&\includegraphics[width=50mm]{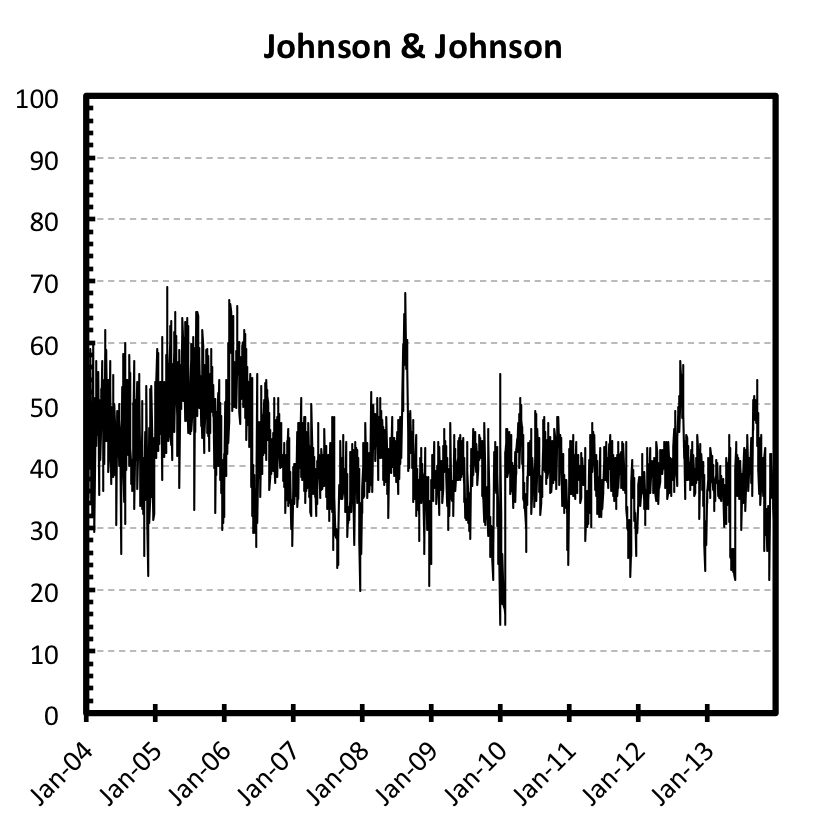}&\includegraphics[width=50mm]{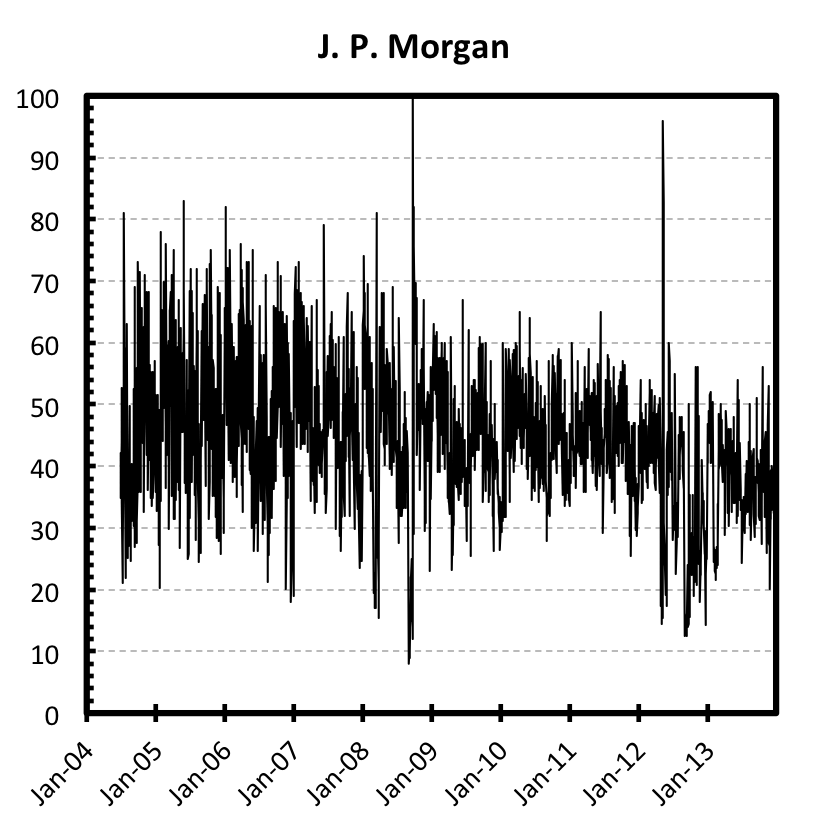}\\
\includegraphics[width=50mm]{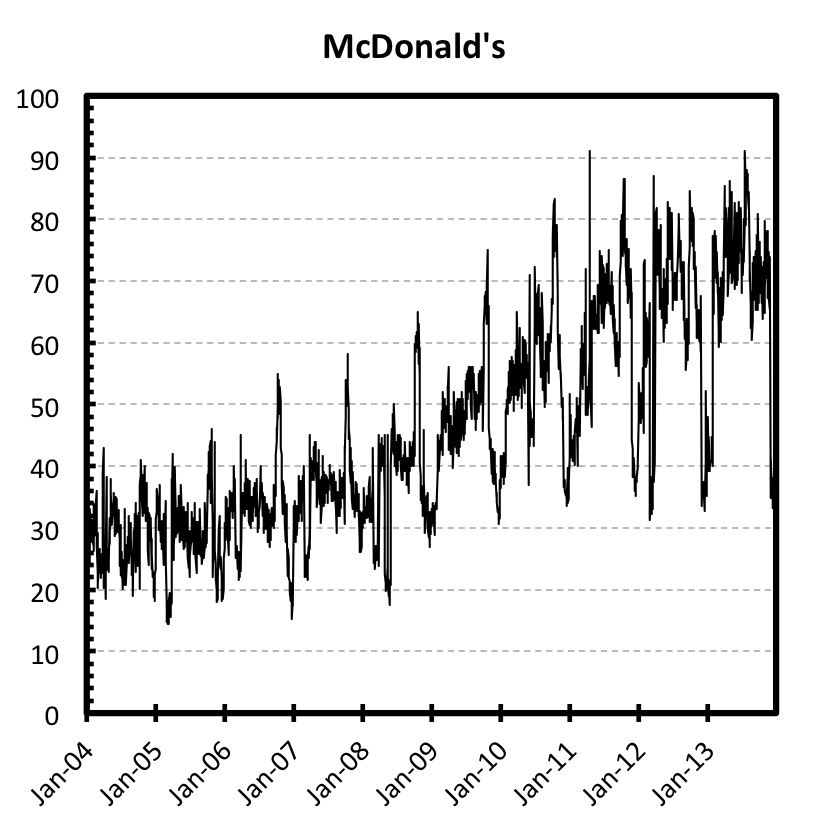}&\includegraphics[width=50mm]{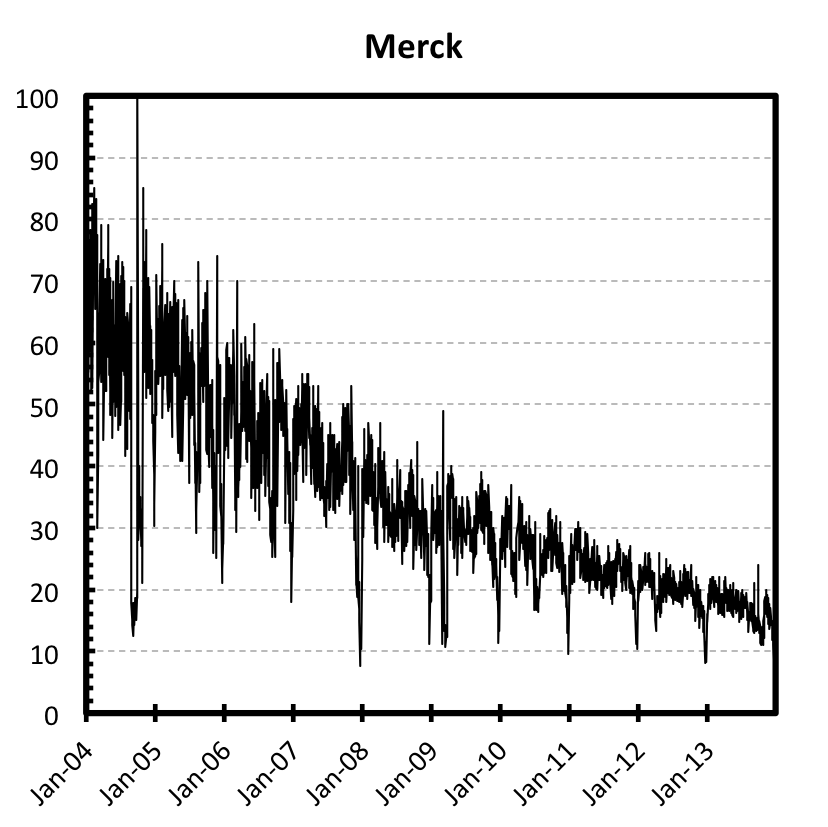}&\includegraphics[width=50mm]{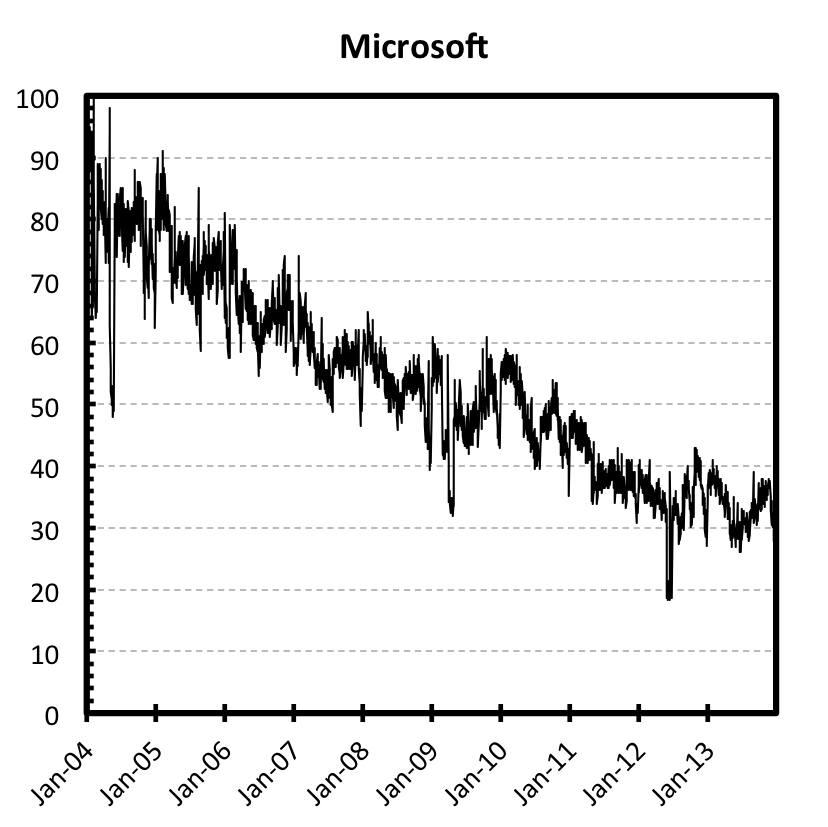}\\
\includegraphics[width=50mm]{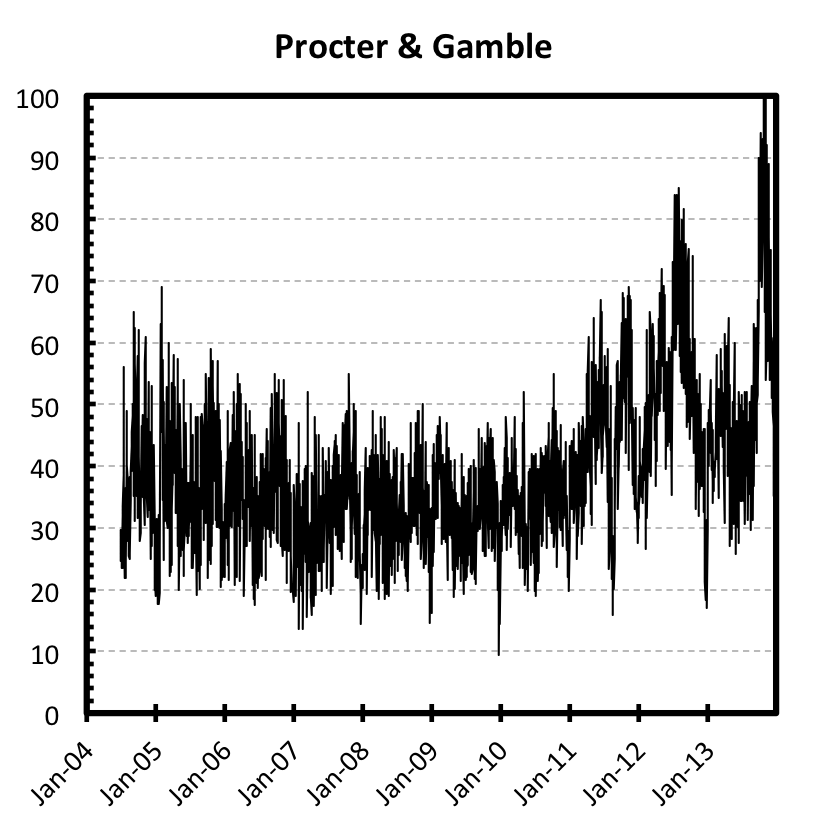}&\includegraphics[width=50mm]{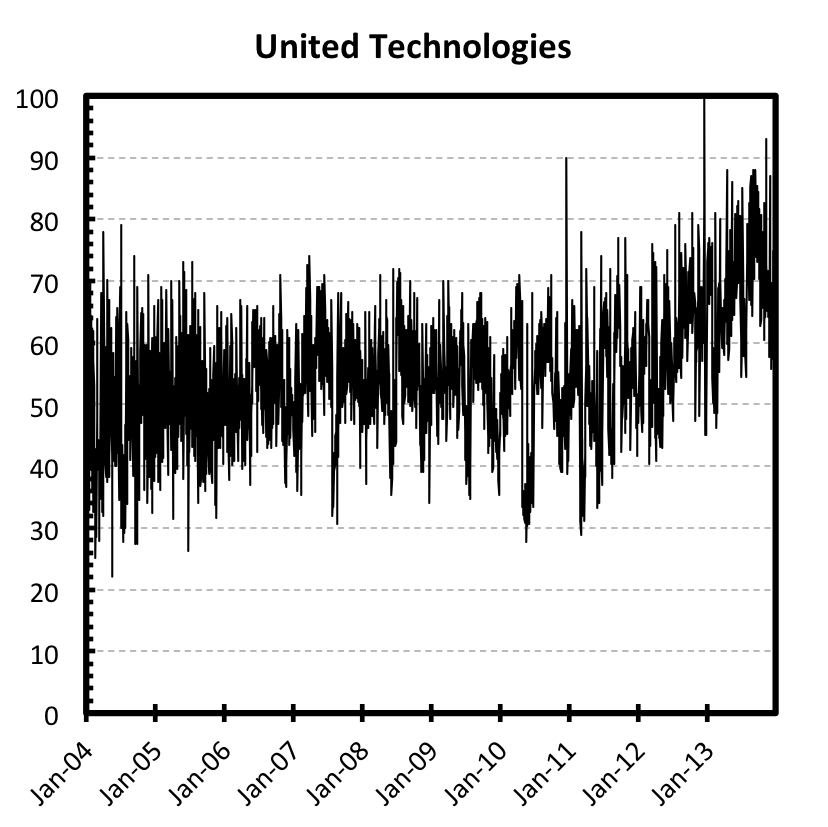}&\includegraphics[width=50mm]{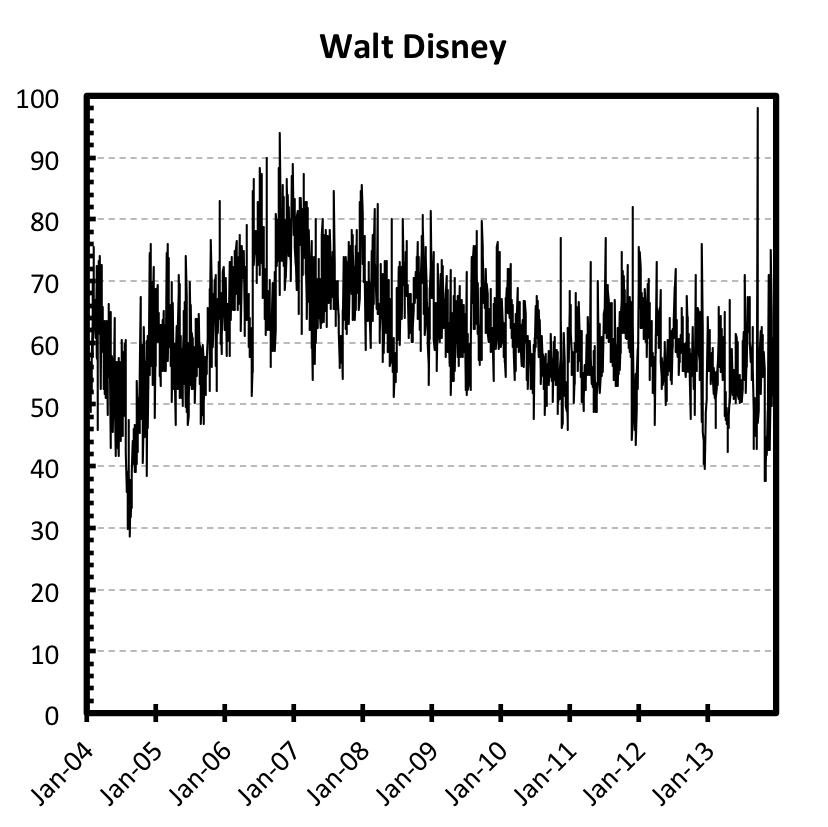}\\
\end{tabular}
\end{center}\vspace{-0.5cm}
\caption{\textbf{Normalized Google searches (Part 2).}\label{fig_series2}
}
\end{figure}

The Google searches for the analyzed stocks are illustrated in Figs. \ref{fig_series1} and \ref{fig_series2}. These uncover that the searching frequencies for the component stocks are very heterogenous. The trends are sharply decreasing (IBM, Merck, Microsoft), slowly decreasing (3M, Boeing, Du Pont, GE, Intel), or reversely increasing rapidly (McDonald's) or slowly (Caterpillar, Coca Cola, Exxon Mobil, Home Depot), or remains quite stable in time (Johnson \& Johnson, J. P. Morgan, United Technologies, Walt Disney). Most of the series show strong seasonal patterns (hence the choice of the DCCA and DMCA techniques which are constructed for such series) mainly connected to the end of the year but also some stronger patterns as for Home Depot. The examined dataset thus provides a complex selection of various dynamic behaviors.

Before we get to the estimated values of Hurst exponent and thus to the type of memory in question, we first test whether the analyzed series are in fact power-law correlated. In Tab. \ref{H}, we present the testing statistics as well as the corresponding $p$-values for the rescaled range and rescaled variance tests as described in the previous section. Apart from two cases (Coca Cola and IBM), the power-law correlations are reported for all series (the null hypothesis is rejected by at least one of the tests at at least 10\% level). It needs to be stressed that levels of the optimal $q$ parameter climb high for all and very high for some cases, sometimes taking into consideration as much as 372 lags of the covariance function (here specifically for IBM). This only strengthens the claim that the analyzed Google series are long-term correlated. This is due to the fact that taking into account already tens of lags practically means considering long-term memory, even more so for hundreds of lags.

\begin{figure}[!htbp]
\begin{center}
\begin{tabular}{ccc}
\includegraphics[width=50mm]{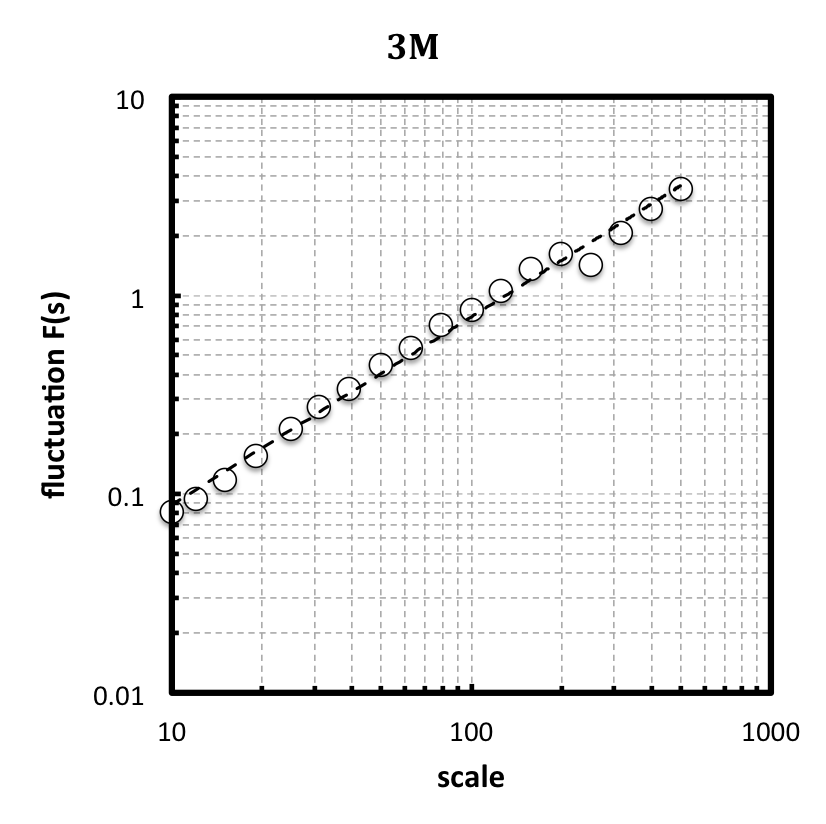}&\includegraphics[width=50mm]{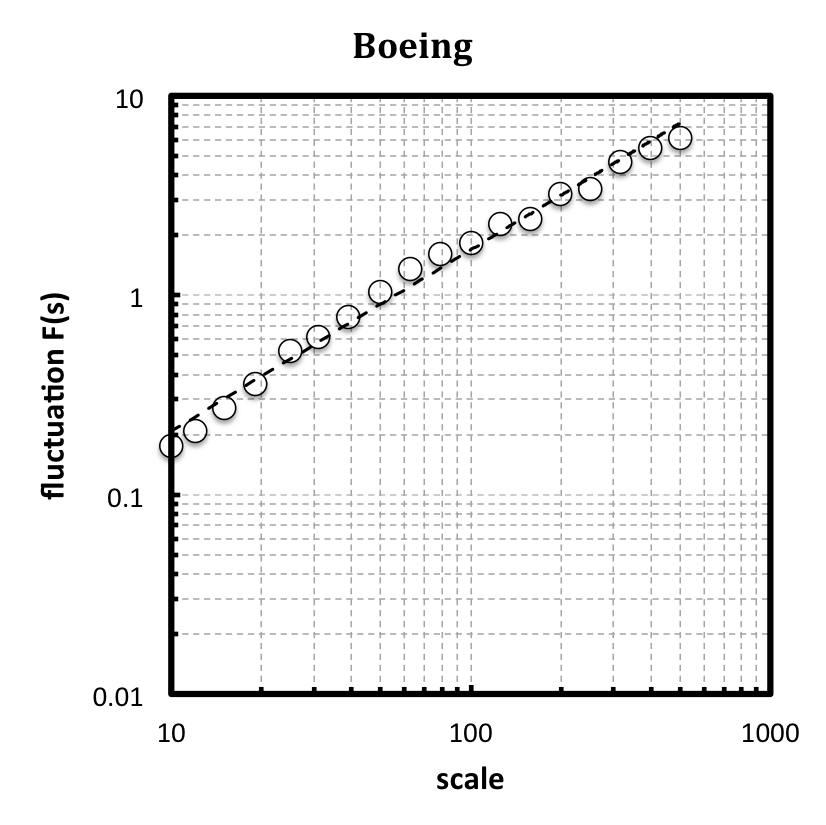}&\includegraphics[width=50mm]{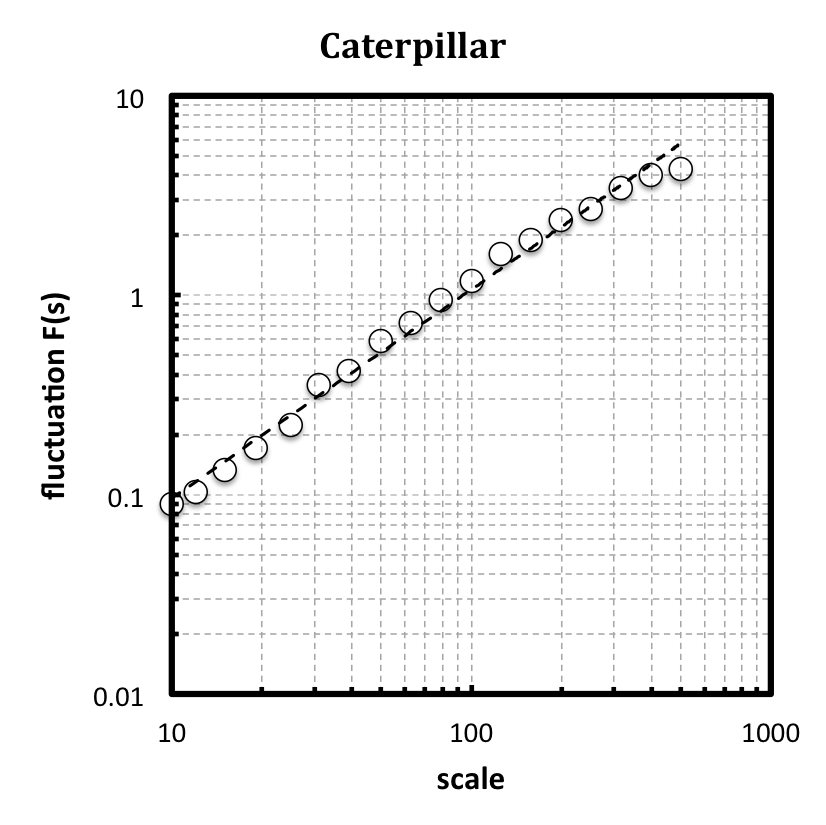}\\
\includegraphics[width=50mm]{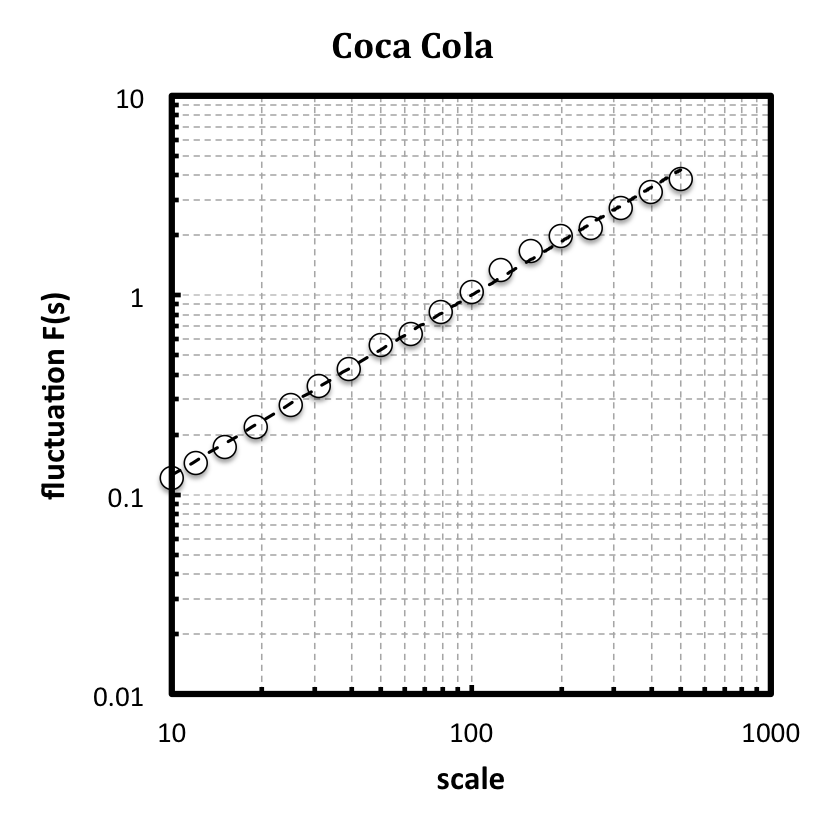}&\includegraphics[width=50mm]{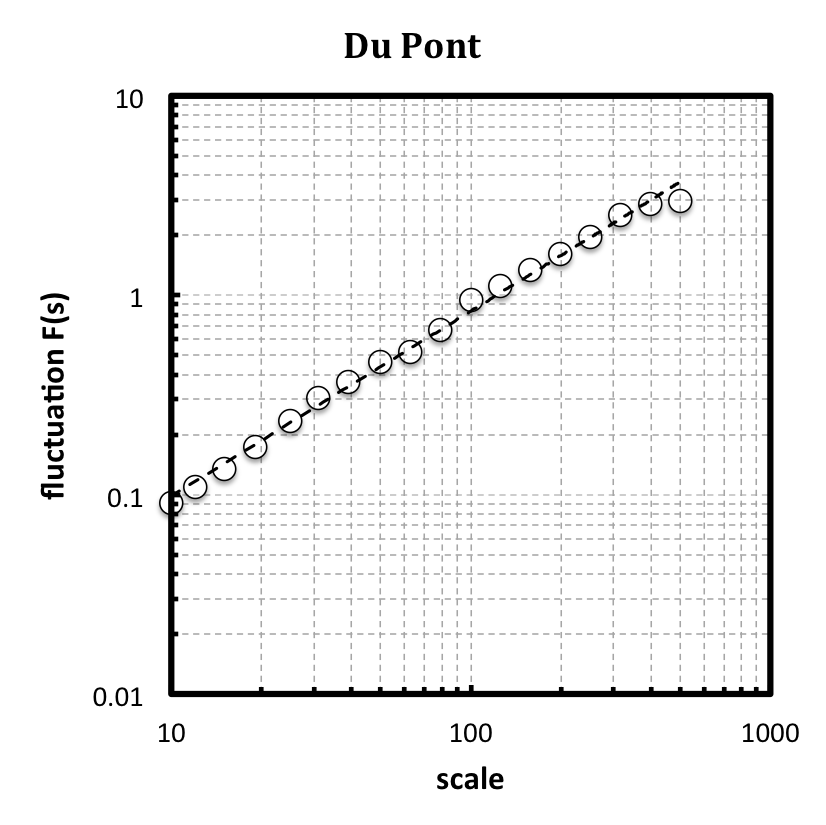}&\includegraphics[width=50mm]{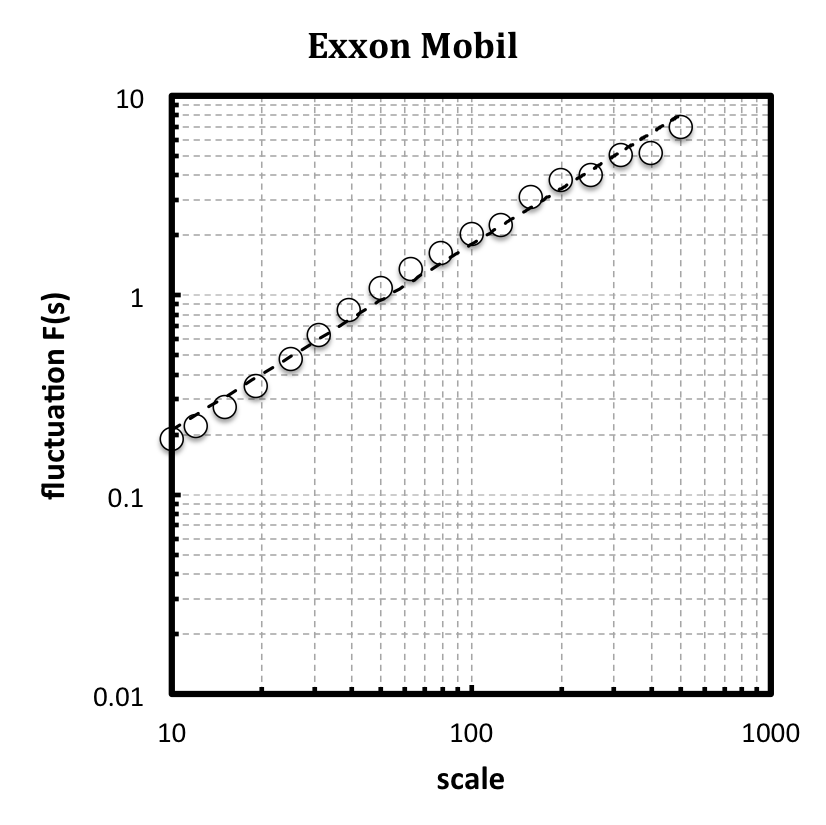}\\
\includegraphics[width=50mm]{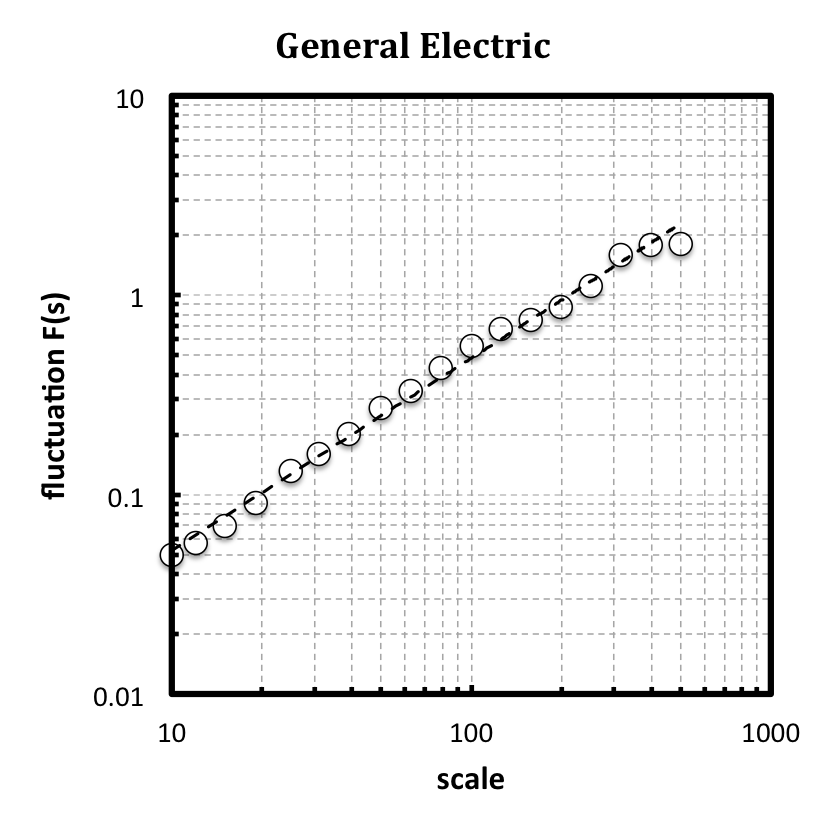}&\includegraphics[width=50mm]{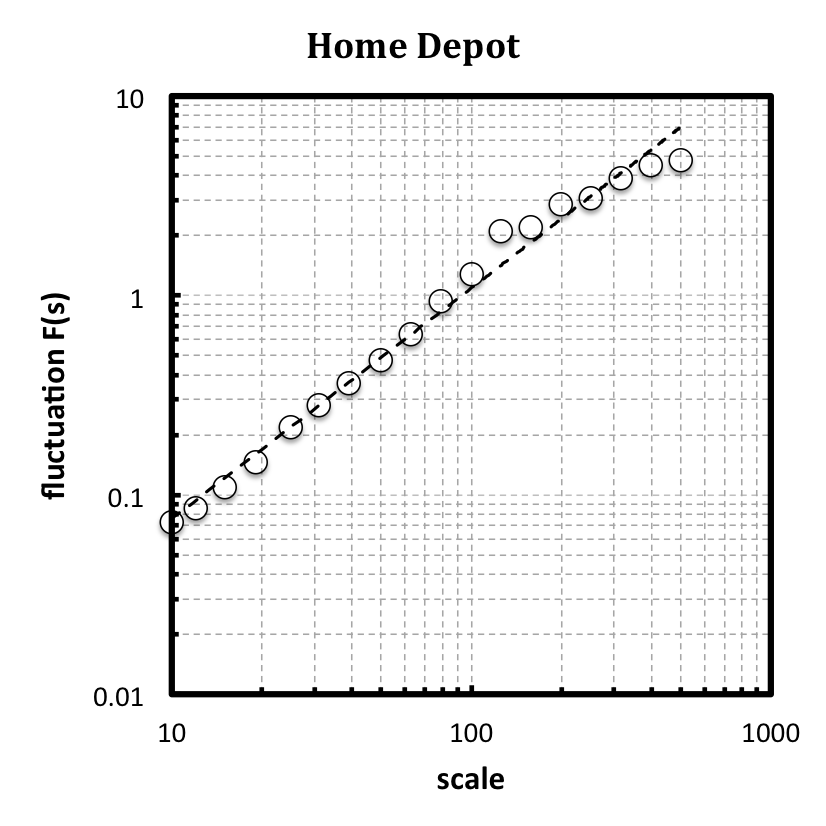}&\includegraphics[width=50mm]{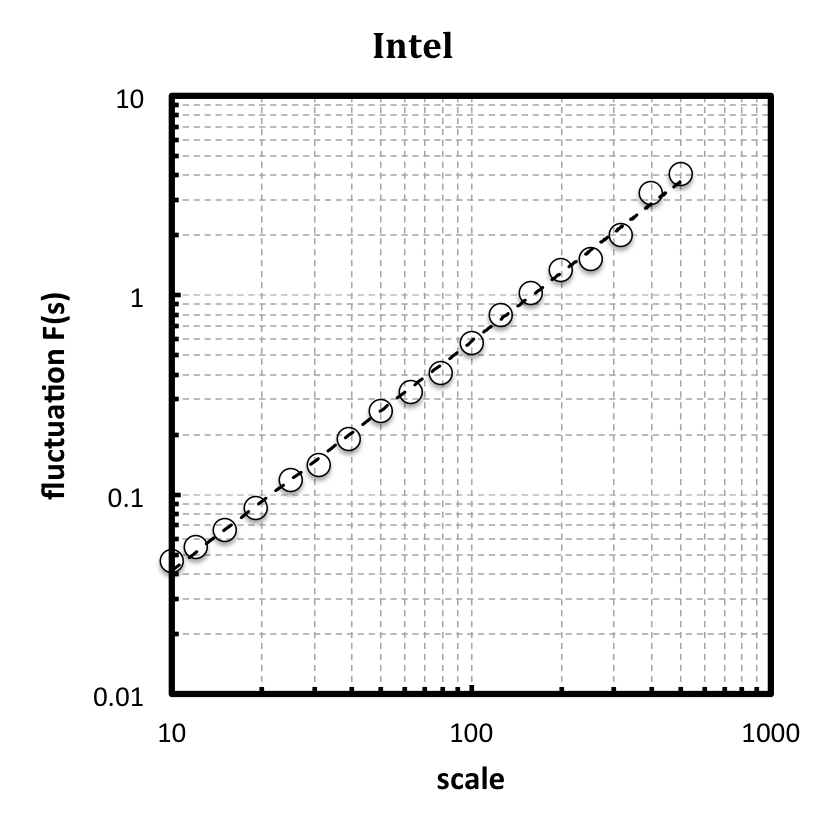}\\
\end{tabular}
\end{center}\vspace{-0.5cm}
\caption{\textbf{DFA scaling of Google searches related to the DJIA component stocks (Part 1).} \footnotesize{Log-log representation shows a profound linear scaling characteristic for long-range correlated processes. Estimated Hurst exponents are summarized in Tab. \ref{H}.}\label{fig_scaling1}
}
\end{figure}

\begin{figure}[!htbp]
\begin{center}
\begin{tabular}{ccc}
\includegraphics[width=50mm]{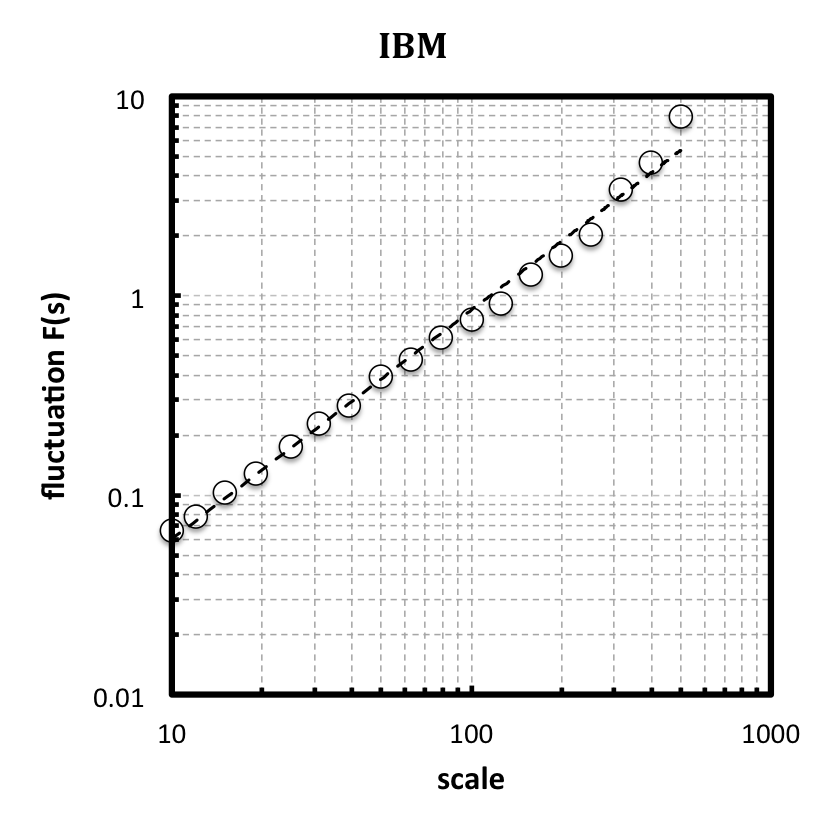}&\includegraphics[width=50mm]{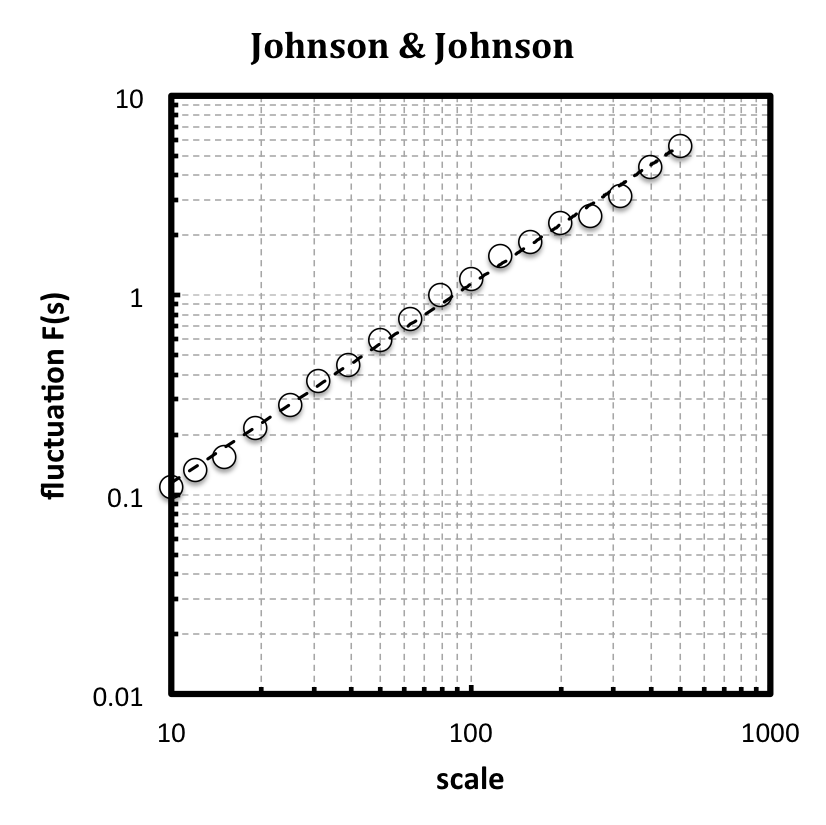}&\includegraphics[width=50mm]{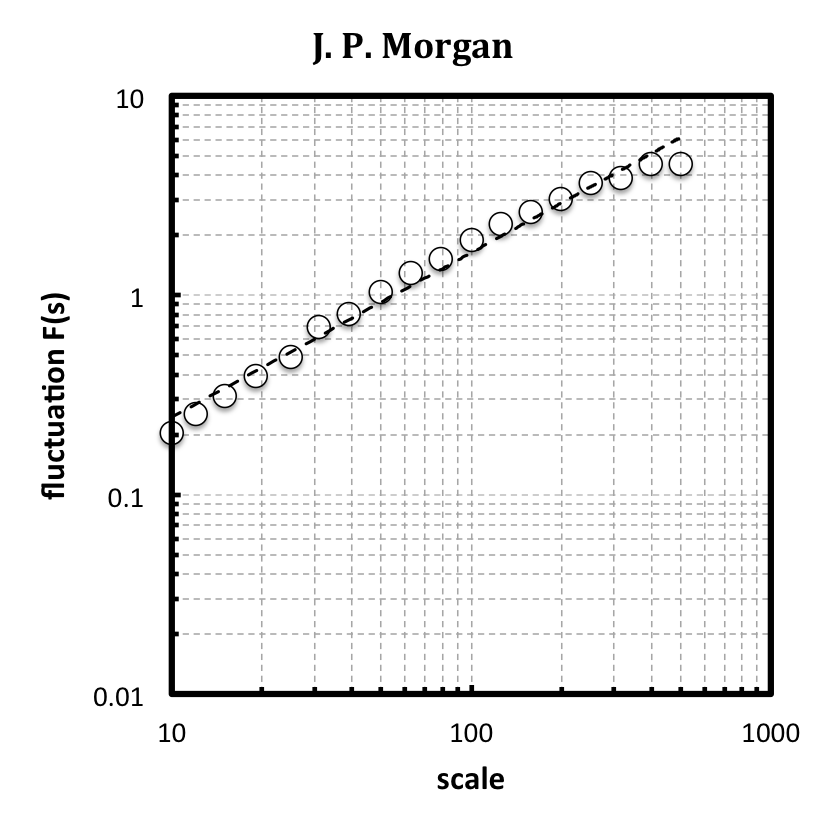}\\
\includegraphics[width=50mm]{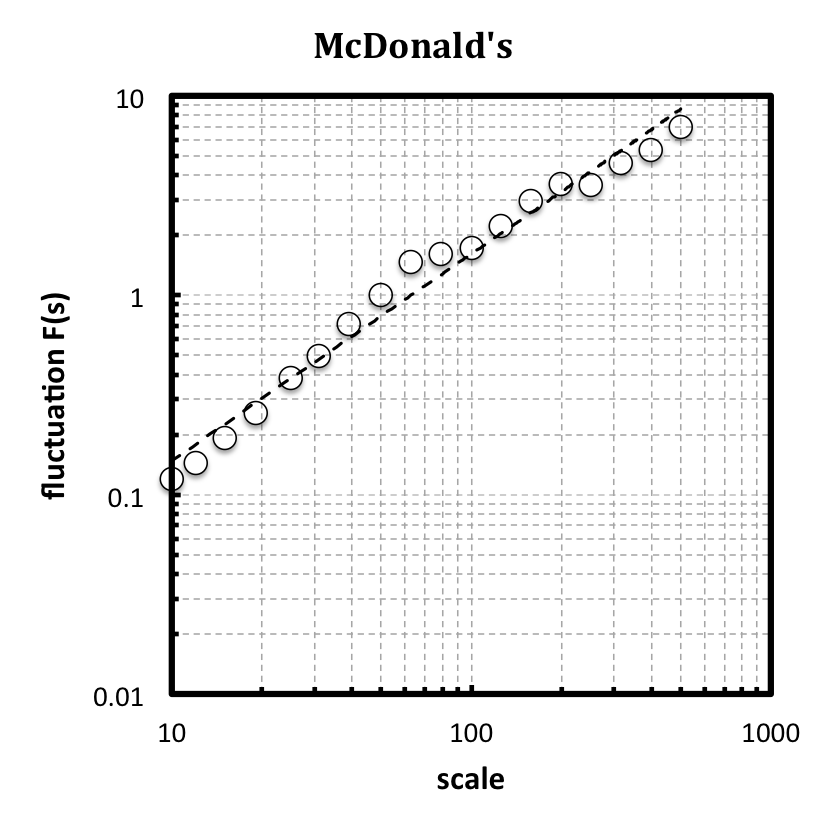}&\includegraphics[width=50mm]{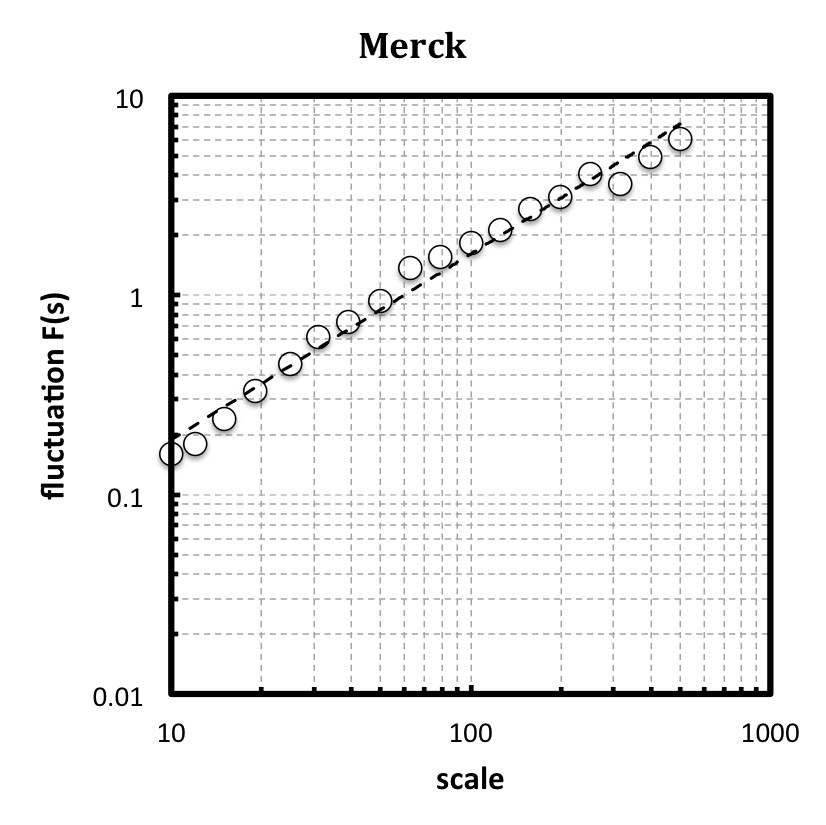}&\includegraphics[width=50mm]{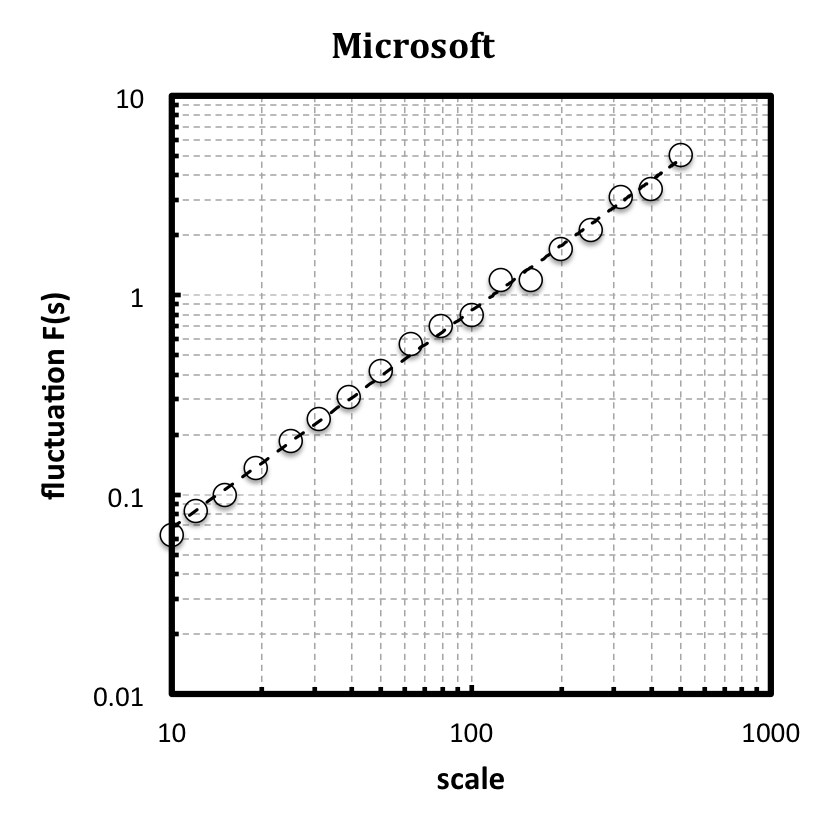}\\
\includegraphics[width=50mm]{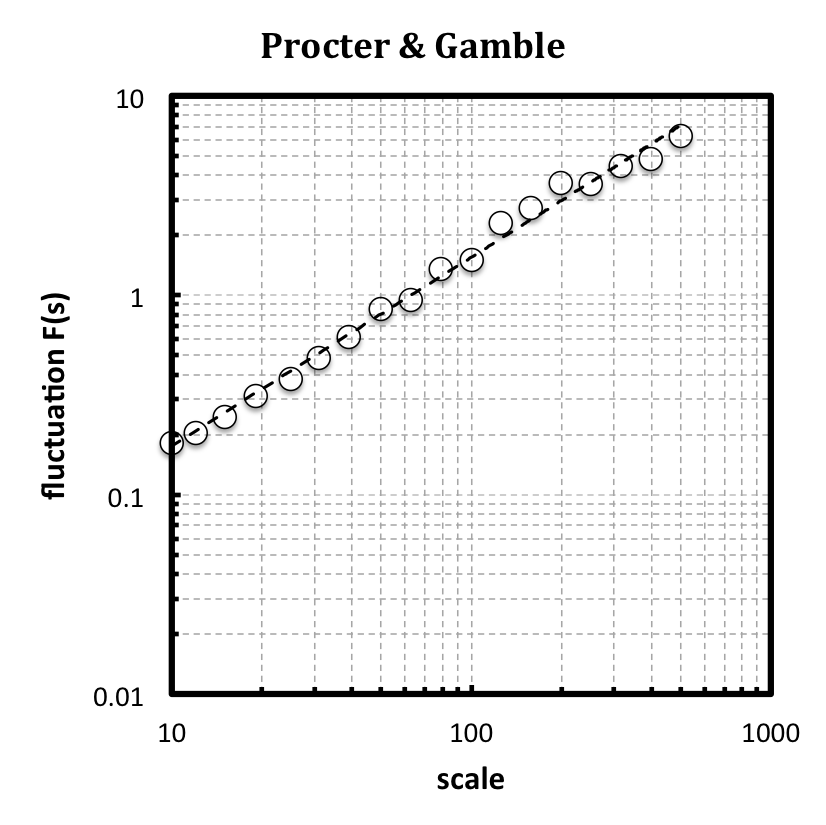}&\includegraphics[width=50mm]{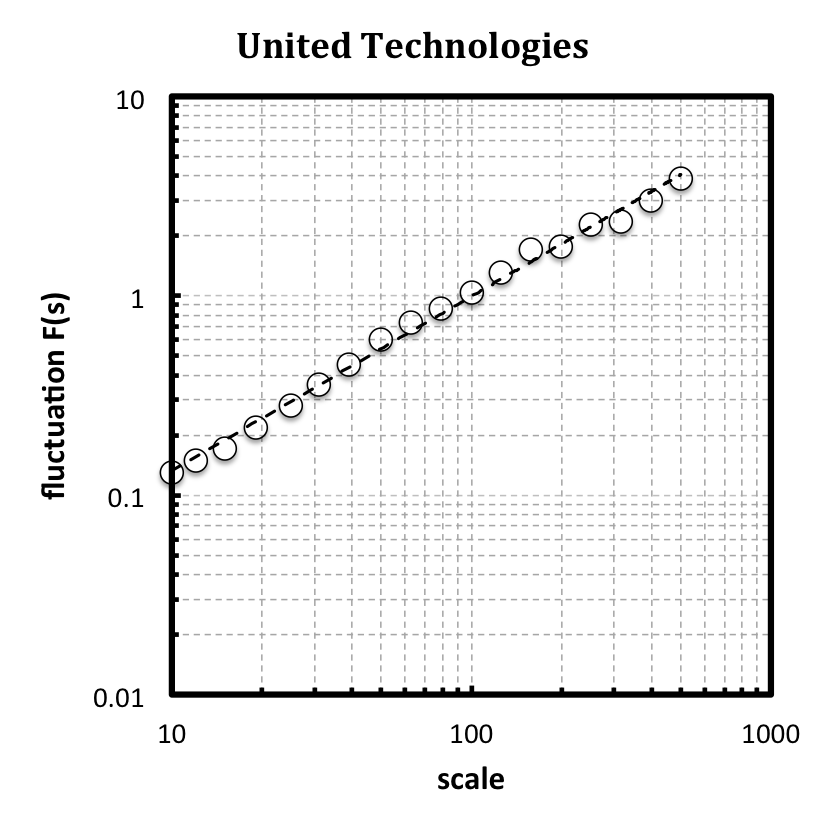}&\includegraphics[width=50mm]{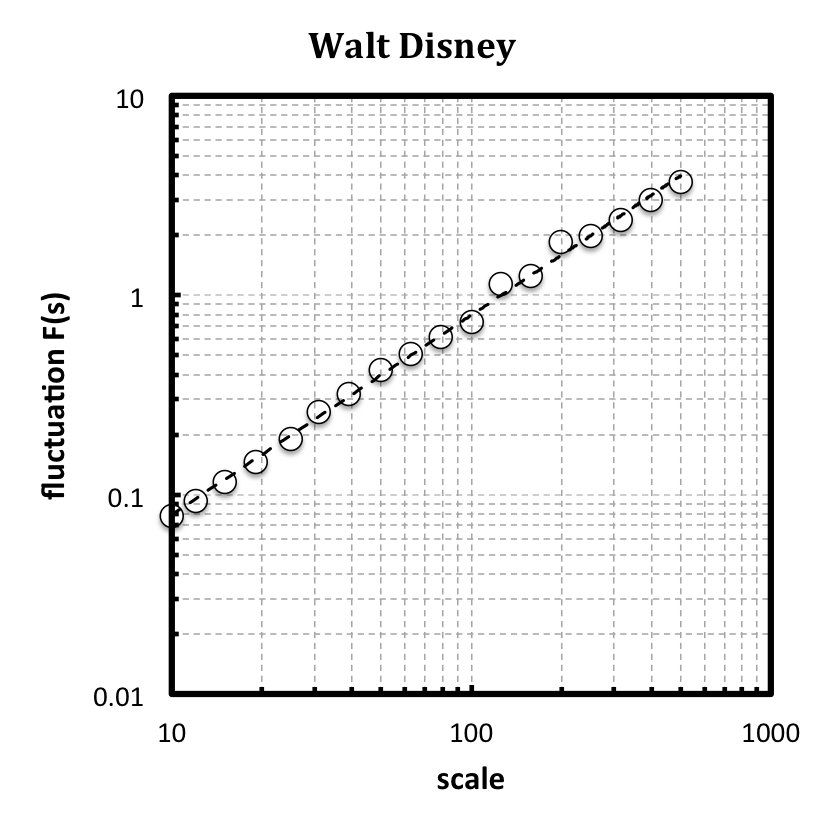}\\
\end{tabular}
\end{center}\vspace{-0.5cm}
\caption{\textbf{DFA scaling of Google searches related to the DJIA component stocks (Part 2).}\label{fig_scaling2}
}
\end{figure}

Tab. \ref{H} also reports the estimated Hurst exponents which are further supported by Figs. \ref{fig_scaling1} and \ref{fig_scaling2}. In the figures, we report an evident power-law scaling of the fluctuation functions according to Eq. \ref{HH}. For all series, the scaling is very stable and the estimated Hurst exponents are thus reliable. Tab. \ref{H} shows that Hurst exponents vary between 0.8 and 1.1. The Google searches are thus strongly persistent for all the analyzed series. Even though the memory is very strong for these series, Hurst exponents still remain below 1.5 which implies that the series stay mean reverting. In the DFA context, this means that even though the online queries series tend to wander away from the long-term trend, they always return to it and they never explode. The fact that the series remain on the edge of stationarity and non-stationarity (around $H=1$) only highlights the need for a careful treatment of such series in multivariate settings which are standardly applied in the empirical literature.

To further illustrate the usefulness of the presented results, we reinvestigate the relationship between Google searches, traded volume and volatility. The traded volume for each component stock of the DJIA index is directly available at finance.yahoo.com as well as are the open, close, high and low prices. We utilize the provided information and construct volatility series using the Garman-Klass variance estimator \cite{Garman1980} defined as
\begin{equation}
\widehat{\sigma}^2_{GK,t}=\frac{(\log(H_t/L_t))^2}{2}-(2\log2-1)(\log(C_t/O_t))^2
\end{equation}
where $H_t$ and $L_t$ are daily highs and lows, respectively, and $C_t$ and $O_t$ are daily closing and opening prices, respectively. The estimator possesses very good statistical properties and serves as an excellent choice without a need of using high-frequency data \cite{Chou2010}. We study a logarithmic transformation of both $\widehat{\sigma}^2_{GK}$ and the traded volume series which is a standard procedure in the applied literature. The transformation of the original variance series allows us to comment on both variance and volatility as the logarithmic variance becomes just twice the logarithmic volatility. 

We examine the correlations between Google searches, traded volume and volatility at various scales using the DCCA and DMCA coefficients. For the DCCA coefficient, we study the correlation between the searches and traded volume, and between the searches and volatility for scales between 10 and 250 with a step of 10. For the DMCA coefficient, we use moving window lengths between 11 and 251 with a step of 10 as well. This way, we obtain comparable results using these methodologies.

\begin{figure}[!htbp]
\begin{center}
\begin{tabular}{cc}
\includegraphics[width=75mm]{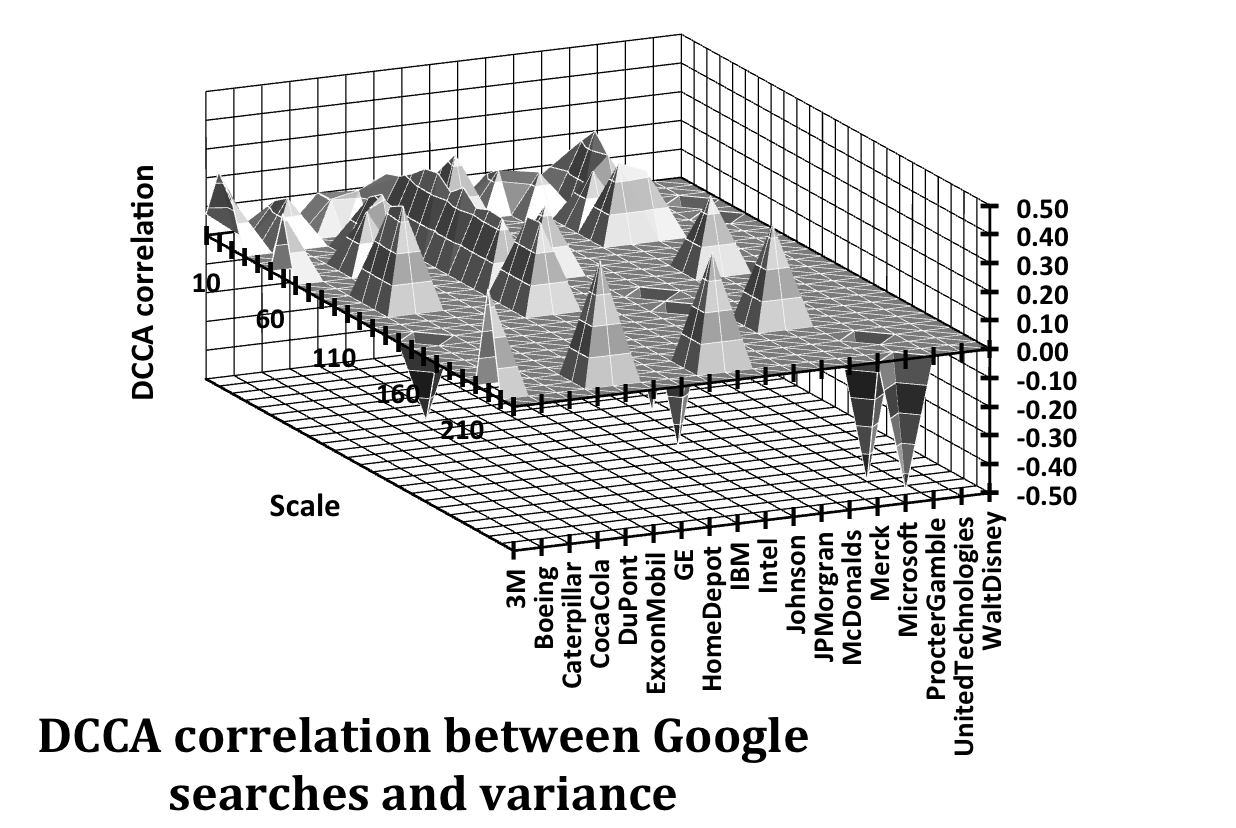}&\includegraphics[width=75mm]{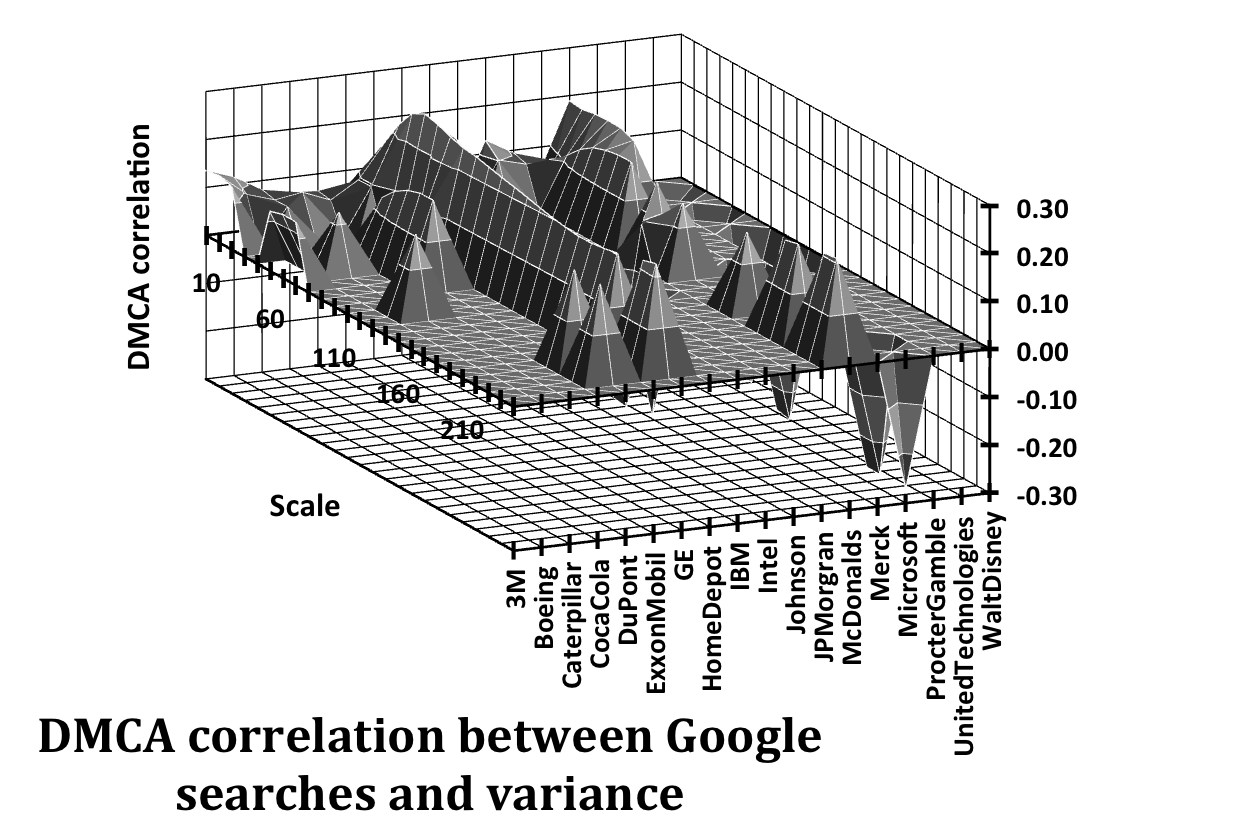}
\end{tabular}
\end{center}\vspace{-0.5cm}
\caption{\textbf{Correlation coefficients between Google searches and variance.} \footnotesize{The correlations are presented for the DCCA (left) and DMCA (right) methods with changing scales and moving average windows, respectively. The results are shown for all analyzed series. Only significant correlations (with $p$-value below 0.1) are reported.}\label{fig1}
}
\end{figure}

\begin{figure}[!htbp]
\begin{center}
\begin{tabular}{cc}
\includegraphics[width=75mm]{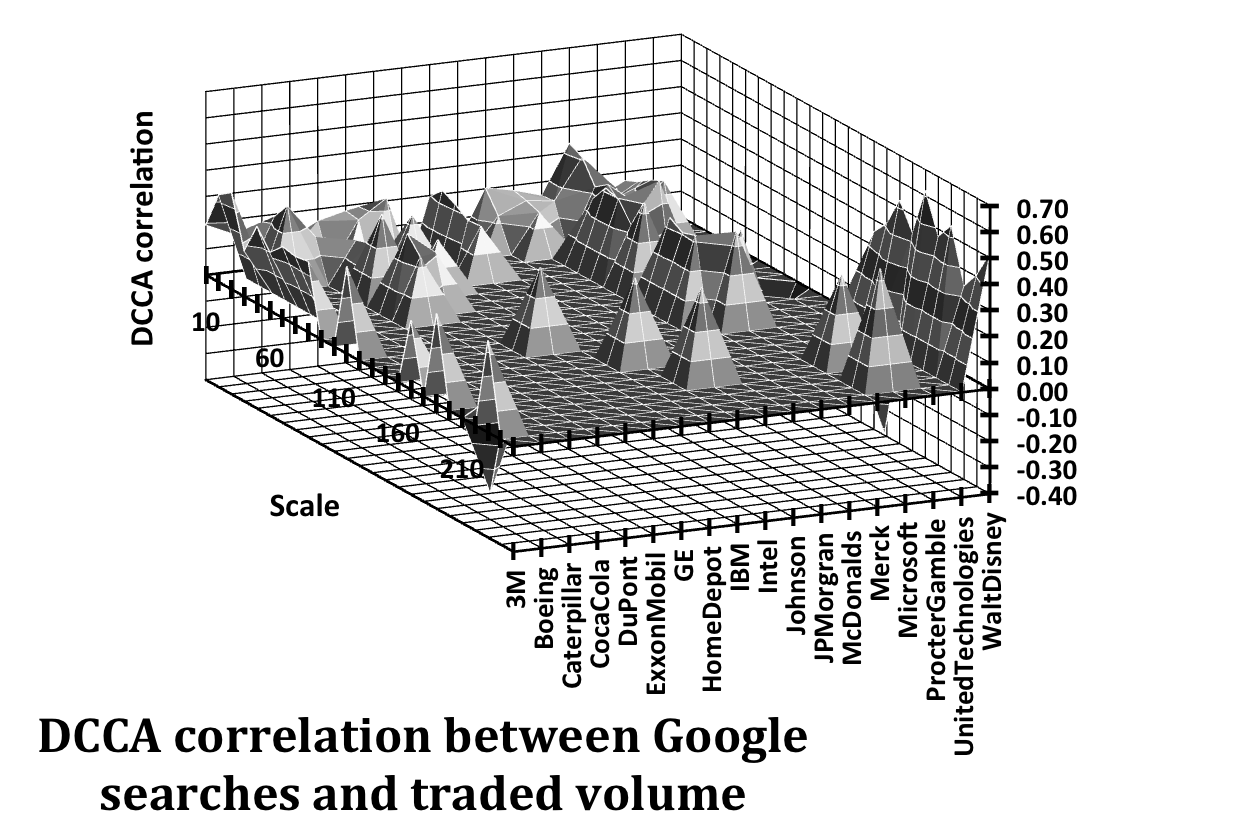}&\includegraphics[width=75mm]{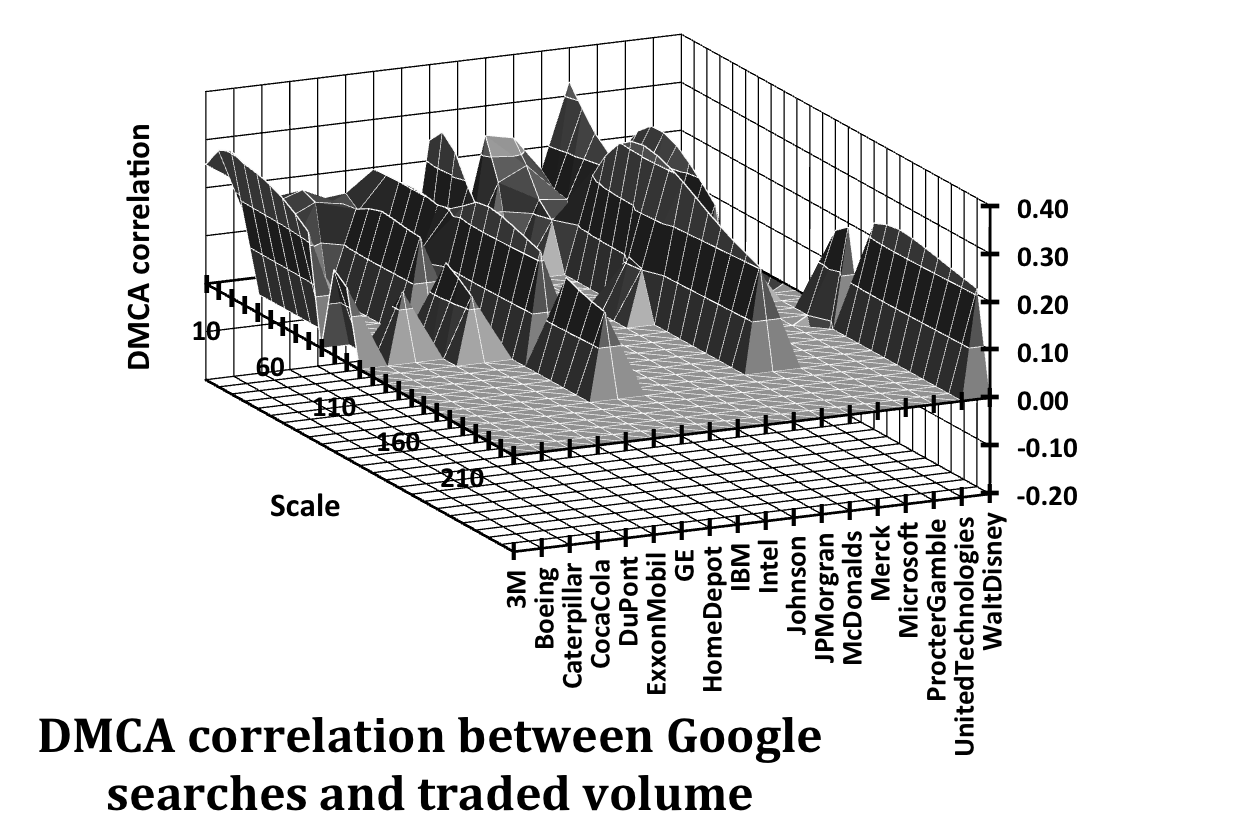}
\end{tabular}
\end{center}\vspace{-0.5cm}
\caption{\textbf{Correlation coefficients between Google searches and traded volume.} \footnotesize{The notation holds from Fig. \ref{fig1}.}\label{fig2}
}
\end{figure}

Figs. \ref{fig1} and \ref{fig2} depict the results for variance and traded volume, respectively, for both methods. Only significant correlations with the $p$-value below 0.10 are reported, the insignificant ones are set to zero. We find several interesting results. First, the DMCA method reports more stable results with more significant coefficients. This is well in hand with the numerical results presented by Kristoufek \cite{Kristoufek2014,Kristoufek2014a}. Second, the correlations for traded volume are in general higher than the ones for volatility. Third, a majority of significant correlations occur at the lower scales. There thus seems to be rather short-term or medium-term relationship between the online searches and the examined financial indicators. In the long-term, only few correlations are identified as significant. And fourth, the level of correlations varies considerably across stock titles. There thus seems to be no universality in the relationship between the searches, and volatility and volume. Tables \ref{rho1} and \ref{rho2} further illustrate the heterogeneity of the results. There, we present the average DCCA and DMCA coefficients across scales together with their significance level. The above mentioned results are supported. First, the significance, level and sign of the correlations vary widely. Second, the DMCA procedure delivers more significant results. And third, the correlations are higher for volume than for volatility. Nevertheless, many of the significant correlations are still below a level of 0.05 and practically all the correlation coefficients fall between -0.2 and 0.2. The correlations are thus very weak even if found statistically significant.

There are still some interesting results mainly connected to various signs of the correlations. For example Microsoft shows some unorthodox behavior for volatility. A positive relationship is usually reported, whereas the search queries for Microsoft are negatively correlated with volatility. Conversely, traded volume shows a positive correlation. It thus seems that general interest in Microsoft is mainly tied with positive news which stabilize the stock price rather than with negative news that would make the price more volatile. Similar dynamics is found for Johnson \& Johnson. The only stock which gives insignificant results for both financial quantities is Merck. Other stocks show either positive and thus expected correlations or only weak negatives ones.

\section{Discussion and conclusions}

We have analyzed the power-law correlations in the online search queries for the DJIA stock components. By reconstructing the daily Google search series, we have been able to obtain enough observations for a valid analysis of long-range dependence. Using the combination of the rescaled range and rescaled variance tests and the detrended fluctuation analysis, we have shown that the online searches are indeed power-law correlated.  Importantly, the level of long-term memory is very high with Hurst exponents around unity (between 0.8 and 1.1) for all the analyzed stocks. Such results suggest that the finance-related online searches have similar dynamic properties to stock variance and traded volume which are themselves power-law correlated. The information flow coming into the stock markets evidently enters the general public interest and keeps it for quite long periods and its dissipation is thus not immediate. The fact that the online searches and implied attention are usually attributed to retail and small investors, such information and attention dissipation fits into the picture of a small investor using the information for decision-making in a longer term. Such persistent dynamics of the series might also arise from indecisiveness of the small investors which would think twice before investing into a specific stock. Online searches then cluster and keep their level for longer time intervals. The results remain fascinatingly universal across the analyzed stocks. Even though the global dynamics of the series is very heterogenous with different speeds of trends or various volatility levels, they all remain strongly persistent with a smooth scaling of fluctuations.

In addition, we have studied the relationship between Google searches, traded volume and volatility using the recently proposed DCCA and DMCA coefficients. The results primarily suggest that there is no universal relationship between the online search queries and the analyzed financial measures. Even though we confirm positive correlation for a majority of pairs, there are several pairs with insignificant or even negative correlations. Further, the correlations vary quite strongly across scales. The online searches have thus retained their potential for financial modeling and various applications but our findings suggest that one needs to carefully study each stock or asset separately as the usefulness of the queries can fluctuate considerably. The reported results do not necessarily contradict some previous findings which find statistically significant connections \cite{Bank2011,Bordino2012,Vlastakis2012,Preis2013} or time varying correlations \cite{Curme2014}. However, we stress that there seems to be no universal and global relationship between the online searches and relevant financial variables (traded volume and volatility).

Our results open an interesting area of further research of the topic. First, the power-law properties of the correlation structure might be observed also in different types of search queries, not necessarily only for stocks or financial markets in general. This would show how information or information seeking dissipates in time and how such behavior connects to other specific phenomena of the relevant time series. Second, the results indicate that the online searches are strongly persistent and on the edge of (non-)stationarity. Such characteristic implies that the simple correlation studies reported in the literature should take this property into consideration as an inappropriate analysis of persistent data using tools for short-range dependent series can produce spurious and in turn misleading results. And third, knowing the basic dynamic properties of the series helps to construct the forecasting models which are of high interest for practitioners, specifically in risk management.

\section*{Acknowledgements}

Support from the Czech Science Foundation under project No. 14-11402P and 
the Research Councils UK via Grant EP/K039830/1 is gratefully acknowledged. I would also like to thank Tomas Vakrman for help with a dataset collection. Google data are registered trademarks of Google Inc., used with permission.


\section*{References}
\bibliography{Google}
\bibliographystyle{unsrt}

\newpage

\begin{table}[!htbp]
\begin{center}
\caption{Searched terms and DJIA component stocks}
\label{query}
\footnotesize
\begin{tabular}{c | c | ccc}
\hline\hline 
No.&Company full name&Company short name&Ticker&Search query\\
\hline
\#1&3M Company&3M&MMM&3M\\
\#2&Caterpillar Incorporated&Caterpillar&CAT&Caterpillar\\
\#3&Coca-Cola Company&Coca Cola&KO&Coca Cola\\
\#4&E. I. du Pont de Nemours and Company&Du Pont&DD&DuPont\\
\#5&Exxon Mobil Company&Exxon Mobil&XOM&Exxon\\
\#6&General Electric Company&General Electric&GE&GE\\
\#7&Home Depot Incorporated&Home Depot&HD&Home Depot\\
\#8&Intel Corporation&Intel&INTC&Intel\\
\#9&International Business Machines&IBM&IBM&IBM\\
\#10&J. P. Morgan Chase&J. P. Morgan&JPM&JP Morgan\\
\#11&Johnson \& Johnson&Johnson \& Johnson&JNJ&Johnson Johnson\\
\#12&McDonald's Corporation&McDonald's&MCD&McDonalds\\
\#13&Merck \& Co., Inc.&Merck&MRK&Merck\\
\#14&Microsoft Corporation&Microsoft&MSFT&Microsoft\\
\#15&Procter \& Gamble Company&Procter \& Gamble&PG&P\&G\\
\#16&The Boeing Company&Boeing&BA&Boeing\\
\#17&United Technologies Corporation&United Technologies&UTX&UTC\\
\#18&Walt Disney Company&Walt Disney&DIS&Disney\\
\hline    \hline
\end{tabular}
\end{center}
\end{table}

\begin{table}[!htbp]
\begin{center}
\caption{Long-term memory tests and estimated Hurst exponent}
\label{H}
\footnotesize
\begin{tabular}{c|c|ccccc|c}
\hline\hline 
Company&\# of obs.&$V_T$&$p$-value&$M_T$&$p$-value&$q^{\ast}$&$H_{DFA}$\\
\hline
3M&2516&2.9267&0.0000&0.0003&0.0000&46&0.9452\\
Boeing&2516&2.7083&0.0000&0.0003&0.0000&65&0.9086\\
Caterpillar&2516&2.0131&0.0120&0.0001&0.0030&33&1.0467\\
Coca Cola&2516&1.6006&0.1848&0.0001&0.2647&23&0.9016\\
Du Pont&2516&2.6857&0.0000&0.0002&0.0000&63&0.9276\\
Exxon Mobil&2265&2.8268&0.0000&0.0003&0.0000&37&0.9313\\
General Electric&2516&2.8344&0.0000&0.0003&0.0000&63&0.9652\\
Home Depot&2516&1.6500&0.0599&0.0001&0.1359&67&1.1557\\
IBM&2516&1.2279&0.6424&0.0001&0.1099&372&1.1478\\
Intel&2516&1.7875&0.0070&0.0001&0.0000&164&1.1486\\
Johnson \& Johnson&2516&2.5457&0.0000&0.0002&0.0000&34&0.9964\\
J. P. Morgan&2393&2.4226&0.0000&0.0002&0.0000&26&0.8261\\
McDonald's&2516&2.0612&0.0000&0.0002&0.0000&127&1.0373\\
Merck&2516&2.3400&0.0000&0.0002&0.0000&88&0.9314\\
Microsoft&2516&1.5279&0.0729&0.0001&0.0060&206&1.0896\\
Procter \& Gamble&2393&3.2807&0.0000&0.0005&0.0000&25&0.9468\\
United Technologies&2516&2.7528&0.0000&0.0002&0.0070&22&0.8747\\
Walt Disney&2516&3.0381&0.0000&0.0004&0.0000&38&1.0003\\
\end{tabular}
\end{center}
\end{table}

\begin{table}[!htbp]
\begin{center}
\caption{Average DCCA and DMCA correlations between volatility and Google searches}
\label{rho1}
\footnotesize
\begin{tabular}{c||ccc|ccc|cc}
\hline\hline 
&$\overline{\rho_{DCCA}}$&$\widehat{\sigma}_{\rho_{DCCA}}$&$p$-value&$\overline{\rho_{DMCA}}$&$\widehat{\sigma}_{\rho_{DMCA}}$&$p$-value&significant&sign\\
\hline
3M&0.0657&0.0270&0.0149&0.0648&0.0112&0.0000&$\checkmark\checkmark$&$+$\\
Boeing&-0.0372&0.0275&0.1771&-0.0281&0.0162&0.0831&$\times\checkmark$&$-$\\
Caterpillar&0.1087&0.0231&0.0000&0.1136&0.0086&0.0000&$\checkmark\checkmark$&$+$\\
Coca Cola&0.0315&0.0214&0.1422&0.0369&0.0093&0.0001&$\times\checkmark$&$+$\\
Du Pont&0.1658&0.0194&0.0000&0.1590&0.0096&0.0000&$\checkmark\checkmark$&$+$\\
Exxon Mobil&0.0432&0.0170&0.0111&0.0411&0.0101&0.0000&$\checkmark\checkmark$&$+$\\
General Electric&0.1827&0.0173&0.0000&0.2181&0.0067&0.0000&$\checkmark\checkmark$&$+$\\
Home Depot&-0.0095&0.0273&0.7283&-0.0193&0.0080&0.0156&$\times\checkmark$&$-$\\
IBM&0.1093&0.0247&0.0000&0.1154&0.0047&0.0000&$\checkmark\checkmark$&$+$\\
Intel&0.0287&0.0252&0.2536&0.0302&0.0125&0.0160&$\times\checkmark$&$+$\\
Johnson \& Johnson&-0.0953&0.0308&0.0020&-0.1431&0.0227&0.0000&$\checkmark\checkmark$&$-$\\
J. P. Morgran&0.0282&0.0176&0.1093&0.0501&0.0094&0.0000&$\times\checkmark$&$+$\\
McDonald's&0.1509&0.0176&0.0000&0.1595&0.0073&0.0000&$\checkmark\checkmark$&$+$\\
Merck&0.0216&0.0328&0.5090&0.0296&0.0228&0.1955&$\times\times$&$0$\\
Microsoft&-0.1635&0.0323&0.0000&-0.1608&0.0250&0.0000&$\checkmark\checkmark$&$-$\\
Procter \& Gamble&-0.0118&0.0172&0.4944&-0.0141&0.0085&0.0960&$\times\checkmark$&$-$\\
United Technologies&-0.0813&0.0259&0.0017&-0.1076&0.0191&0.0000&$\checkmark\checkmark$&$-$\\
Walt Disney&0.0841&0.0195&0.0000&0.0509&0.0052&0.0000&$\checkmark\checkmark$&$+$\\
\hline    \hline
\end{tabular}
\end{center}
\end{table}

\begin{table}[!htbp]
\begin{center}
\caption{Average DCCA and DMCA correlations between traded volume and Google searches}
\label{rho2}
\footnotesize
\begin{tabular}{c||ccc|ccc|cc}
\hline\hline 
&$\overline{\rho_{DCCA}}$&$\widehat{\sigma}_{\rho_{DCCA}}$&$p$-value&$\overline{\rho_{DMCA}}$&$\widehat{\sigma}_{\rho_{DMCA}}$&$p$-value&significant&sign\\
\hline
3M&0.1599&0.0302&0.0000&0.1667&0.0179&0.0000&$\checkmark\checkmark$&$+$\\
Boeing&-0.0280&0.0251&0.2647&-0.0468&0.0186&0.0117&$\times\checkmark$&$-$\\
Caterpillar&0.0785&0.0266&0.0032&0.0848&0.0144&0.0000&$\checkmark\checkmark$&$+$\\
Coca Cola&0.0331&0.0230&0.1508&0.0436&0.0135&0.0013&$\times\checkmark$&$+$\\
Du Pont&0.1181&0.0184&0.0000&0.1051&0.0148&0.0000&$\checkmark\checkmark$&$+$\\
Exxon Mobil&0.0936&0.0183&0.0000&0.0889&0.0073&0.0000&$\checkmark\checkmark$&$+$\\
General Electric&0.1093&0.0218&0.0000&0.1994&0.0056&0.0000&$\checkmark\checkmark$&$+$\\
Home Depot&0.0262&0.0341&0.4423&0.0382&0.0274&0.1638&$\times\times$&$0$\\
IBM&0.1547&0.0151&0.0000&0.1407&0.0093&0.0000&$\checkmark\checkmark$&$+$\\
Intel&0.0357&0.0254&0.1609&0.0345&0.0096&0.0003&$\times\checkmark$&$+$\\
Johnson \& Johnson&0.0211&0.0254&0.4066&-0.0172&0.0173&0.3192&$\times\times$&$0$\\
J. P. Morgran&0.1258&0.0195&0.0000&0.1008&0.0077&0.0000&$\checkmark\checkmark$&$+$\\
McDonald's&0.2032&0.0185&0.0000&0.2168&0.0133&0.0000&$\checkmark\checkmark$&$+$\\
Merck&0.0463&0.0447&0.3009&0.0328&0.0318&0.3022&$\times\times$&$0$\\
Microsoft&0.0993&0.0248&0.0001&0.0937&0.0087&0.0000&$\checkmark\checkmark$&$+$\\
Procter \& Gamble&0.0724&0.0212&0.0007&0.0865&0.0199&0.0000&$\checkmark\checkmark$&$+$\\
United Technologies&-0.0856&0.0238&0.0003&-0.0809&0.0156&0.0000&$\checkmark\checkmark$&$-$\\
Walt Disney&0.2297&0.0437&0.0000&0.1317&0.0209&0.0000&$\checkmark\checkmark$&$+$\\
\hline    \hline
\end{tabular}
\end{center}
\end{table}

\newpage

\section*{Figures and tables}

\end{document}